\documentclass[%
 reprint,
nofootinbib,
 amsmath,amssymb,
 aps,
prstab,
floatfix,
]{revtex4-1}

\usepackage{url}
\usepackage{graphicx}
\usepackage{dcolumn}
\usepackage{bm}

\begin{document}

\title{Active Suppression of Microphonics Detuning in high $Q_L$ Cavities}

\author{Nilanjan Banerjee} \email{nb522@cornell.edu}
\author{Georg Hoffstaetter}\author{Matthias Liepe}\author{Peter Quigley}\author{Zeyu Zhou}
 \affiliation{Department of Physics, Cornell University, Ithaca, New York 14853, USA}

\date{\today}
\begin{abstract}
Operation of Superconducting Radio Frequency (SRF) cavities with high loaded quality factors is becoming increasingly preferred for applications which involve low beam loading including Energy Recovery Linacs (ERL). Vibration induced microphonics detuning poses a major operational bottleneck in these low bandwidth systems, adversely affecting field stability. Besides passive measures of mitigating the vibration sources, modern SRF cavities are also attached to fast tuners incorporating piezo-electric actuators. We demonstrate the narrow band active noise control algorithm for realizing active resonance control and propose a modification based on the Least Mean Square approach to adaptively tune the control parameters and study it's stability and performance. We discuss our experience of using passive mitigation techniques while commissioning the Main Linac Cryomodule of the Cornell-BNL ERL Test Accelerator (CBETA) and report a net reduction in peak detuning by more than a factor of 2 in its unstiffened cavities. Finally, we demonstrate stable performance of our resonance control system with consistent reduction of peak microphonics detuning by almost a factor of 2 on multiple cavities.
\end{abstract}

\maketitle

\section{Introduction}
Modern particle accelerators are reaching the pinnacle of efficiency using Superconducting Radio Frequency (SRF) cavities which are characterized by low thermal loses arising from high intrinsic quality factors ($Q_0\gtrsim 10^{10}$) \cite{Hasan2014}. The microwave power requirements of such SRF cavities depend on the effective beam loading and the loaded quality factor $Q_L$ used in operation. In situations of high beam loading, they are operated with a comparatively low $Q_L$ in order to couple the required power into the beam, such as in the LHC \cite{Mastorides2010}, CESR \cite{Belomestnykh2001}, NSLS-II \cite{Rose2011} and many others. However, in new applications such as light source Linacs (eg. LCLS-II \cite{Doolittle2015}, XFEL \cite{Branlard2013}) and in Energy Recovery Linacs (eg. CBETA \cite{Hoffstaetter2017}, bERLinPro \cite{AboBakr2018}), high $Q_L$ are becoming common due to the low or negligible beam loading involved. Low beam loading implies the reduction of the RF power requirements and allows the use of efficient solid state amplifiers.

However, the limited bandwidth arising from large $Q_L$ make RF systems more sensitive to detuning when operating at a fixed frequency, as during linac operation. Transient changes in the resonant frequency of the cavity resulting from mechanical deformations change its response to the microwaves coming through the fundamental power coupler. Due to enhanced reflection of the incoming waves from a detuned cavity, more power is needed to maintain a stable field. The interaction of the field with the wall currents is one mechanism leading to mechanical deformation and is known as Lorentz Force Detuning (LFD). This leads to transient detuning as a function of the field inside the cavity and is important for pulsed RF systems. Vibrations inside cryomodules couple into the cavity walls causing transient deformations in its shape resulting in microphonics detuning. The RF power $P$ consumed by a detuned cavity to maintain a voltage $V$ with zero beam loading is given by \cite{Liepe2003},
\begin{equation}
    P = \frac{V^2}{8\frac{R}{Q}Q_L} \frac{\beta + 1}{\beta} \bigg[ 1 + \bigg( \frac{2 Q_L\Delta \omega}{\omega_0}\bigg)^2 \bigg]
\end{equation}
Where $Q_L$ is the loaded quality factor, $\beta$ is the coupling factor, $R/Q$ is the shunt impedance in circuit definition and $\Delta \omega$ is the detuning of the SRF cavity. Hence, the maximum voltage which can be stably sustained in a cavity depends on the peak microphonics detuning and is constrained by the peak forward power available from the amplifiers.

\begin{figure}
    \centering
    \includegraphics[scale = 0.2]{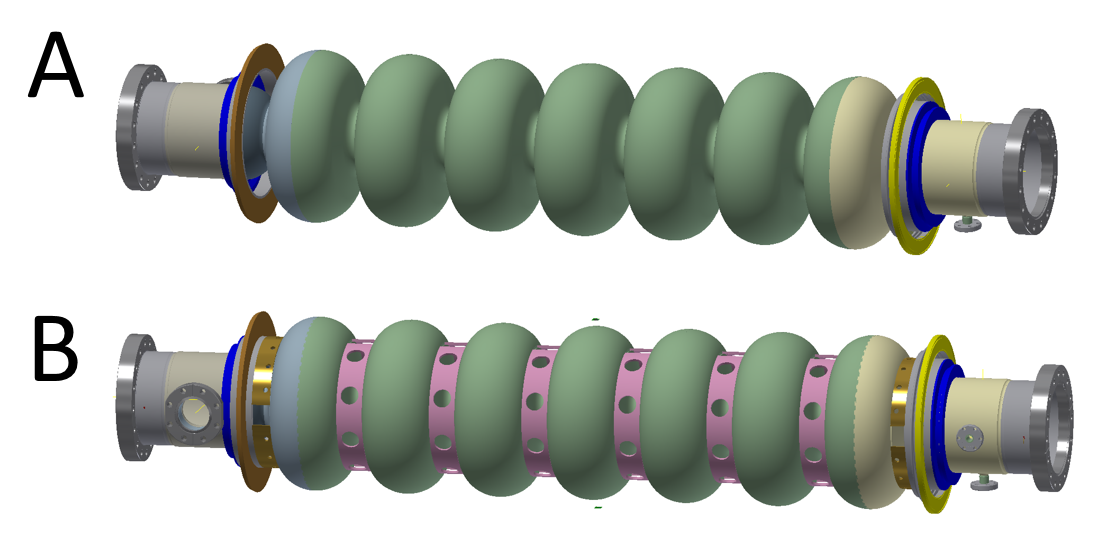}
    \caption{Main linac cavities used in the CBETA project. A is an unstiffened cavity, while B has stiffening rings welded on to it to make it less susceptible to pressure changes in the Helium bath.}
    \label{fig:mlccav}
\end{figure}
Suppression of peak detuning is important in machines operating with high $Q_L$ and designing cavities mechanically less sensitive to vibrations is one way of achieving this goal. Cavities fabricated with metal rings welded on to them as shown in Fig.~\ref{fig:mlccav} can be designed to be less sensitive to vibrations. Depending on whether the machine will be pulsed or CW, the shape and location of the stiffening rings may be optimized to reduce the effect of LFD or increase its stiffness towards external forces respectively. \cite{Posen2012PRAB} In this paper, we discuss suppression of the vibration sources and describe active compensation of microphonics detuning to reduce peak power consumption.

In the next section, we describe the design and operation of fast tuners while modelling them as a linear time invariant system and further explore their non-linear behavior. Using the linear model, we develop a Least Mean Square (LMS) control system based on narrow band Active Noise Control to command the piezo-electric actuators and analyze its performance and stability. Next, we catalog the microphonics sources we found during the commissioning of the Main Linac Cryomodule (MLC) used in CBETA and the measures we took to mitigate them. We then report on the results of using the active control algorithm during RF operations. Finally, we present a summary of our work and propose some improvements to the resonance control system.

\section{Fast Tuner}
\begin{figure}
    \centering
    \includegraphics[scale=0.7]{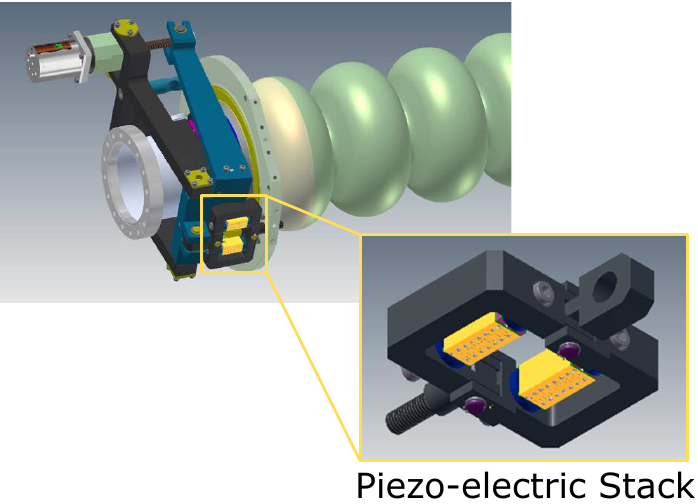}
    \caption{Tuner used for the main linac in the CBETA project based on the Saclay - I design with added fast actuators. The two piezo-electric actuator stacks in yellow provides fast tuning capabilities to the tuner.}
    \label{fig:fasttuner}
\end{figure}

Mitigation of vibration sources is the preferred method of suppressing microphonics, however an active resonance control mechanism is equally important. By further reducing peak detuning, it improves the margin of power consumption with respect to the maximum capability of the microwave amplifier. It also provides an emergency mitigation mechanism against new sources of microphonics until they are found and suppressed. Active control of microphonics requires the use of fast tuners with acoustic response time scales such as the one shown in Fig.~\ref{fig:fasttuner}. \cite{Eichhorn2014} While a stepper motor drives the slow movement of the tuner over a large range, the piezo-electric actuators drive fast movement with a range of 2 kHz \cite{Posen2012ipac} which is almost 100 times the operating bandwidth of the cavity. The response of the cavity resonance frequency to voltages applied to the actuator greatly influences the design of the active resonance control system.

\subsection{Linear Response}

\begin{figure}
    \centering
    \includegraphics[scale=0.5]{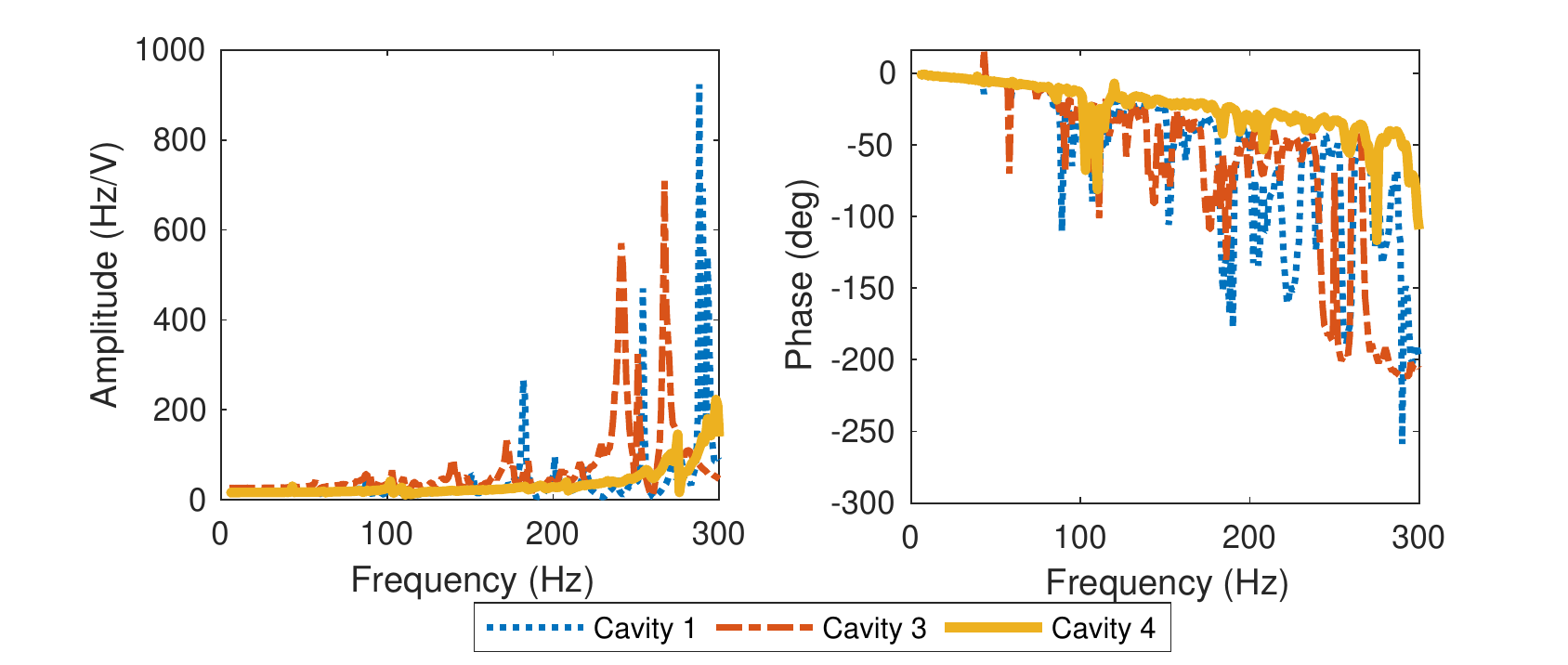}
    \caption{Tuner response amplitude and phase as functions of excitation frequency for three cavities of the main linac in the CBETA project. The plots show multiple strong resonances above 200 Hz for unstiffened cavities 1 and 3. Cavity 4 is fitted with stiffening rings which suppress the low frequency eigenmodes of the structure.}
    \label{fig:tunerresp}
\end{figure}

A Linear Time Invariant (LTI) response is a simple model which can predict the output of a linear system to a given excitation. In the case of the fast tuner, the forces generated by the piezo-electric crystals are approximately a linear function of the voltage applied to it. As long as the materials are in the elastic regime, the strain of different locations of the cavity driven by these forces is governed by a linear partial differential equation. This implies that the detuning of the cavity may be written as a linear ordinary differential equation with the piezo voltage as its source term. This motivates the use of linear response theory which models the output of the system in time domain as a convolution of the input with an impulse response function as follows,
\begin{equation}
    \delta f_{\rm tuner} (t) = \int_0^t \tau(t - t') u_{\rm pz}(t') \rm{d}t' 
    \label{eq:impresp}
\end{equation}
where $\delta f_{\rm tuner}$ is the change in resonant frequency of the cavity generated by the tuner, $u_{\rm pz}$ is the voltage applied and $\tau(t)$ is the impulse response function which encodes the mechanics of the tuner. The output linearly depends on the input since if we scale and add inputs together, the output can be similarly obtained by scaling and adding the original outputs. The response is also time invariant, i.e. if we shift the input in time, the output also suffers the same shift. We can use this LTI model to design the active control system.

We can measure the linear response of the tuner based on the LTI model in two ways. We can directly determine the impulse response by measuring the detuning as a function of time right after applying a short (delta function) voltage pulse to the piezo-electric actuator. However measuring the response function in the time domain is complicated by the presence of strong microphonics and a frequency domain description lends well to the measurement. Applying the Fourier transform on both sides of Eq.~(\ref{eq:impresp}) we obtain,
\begin{equation}
    \delta \tilde{f}_{\rm tuner}(\omega) = \tau(\omega) \tilde{u}_{\rm pz} (\omega)
\end{equation}
where $\delta \tilde{f}_{\rm tuner}(\omega)$ and $\tilde{u}_{\rm pz} (\omega)$ are the Fourier transforms of detuning and voltage respectively. $\tau(\omega)$ is the frequency domain tuner transfer function which encodes both the amplitude of the response and the phase shift generated by the tuner. We determine the frequency response by exciting a sine wave on the actuator with different amplitudes and varying frequency while measuring the amplitude and phase of the resultant sine wave of detuning.

The transfer functions measured on three cavities of the main linac in CBETA are shown in Fig.~\ref{fig:tunerresp}. All the transfer functions show a region of flat amplitude and linear phase response in the range of low frequencies up to 30 Hz; this makes the use of simple algorithms like proportional integral control feasible for attenuating low frequency microphonics. The large peaks in amplitude correspond to resonances and they are accompanied by large swings in the phase response of the tuner; this limits feedback control at these frequencies. The measurements also verify one of the design goals of stiffening cavities, shifting the lowest mechanical eigenmode to a higher frequency. The transfer function data can be used to construct an LTI model of the tuner and is used to analyze the stability of the control algorithm used for resonance control.

\begin{figure}
    \centering
    \includegraphics[scale = 0.34]{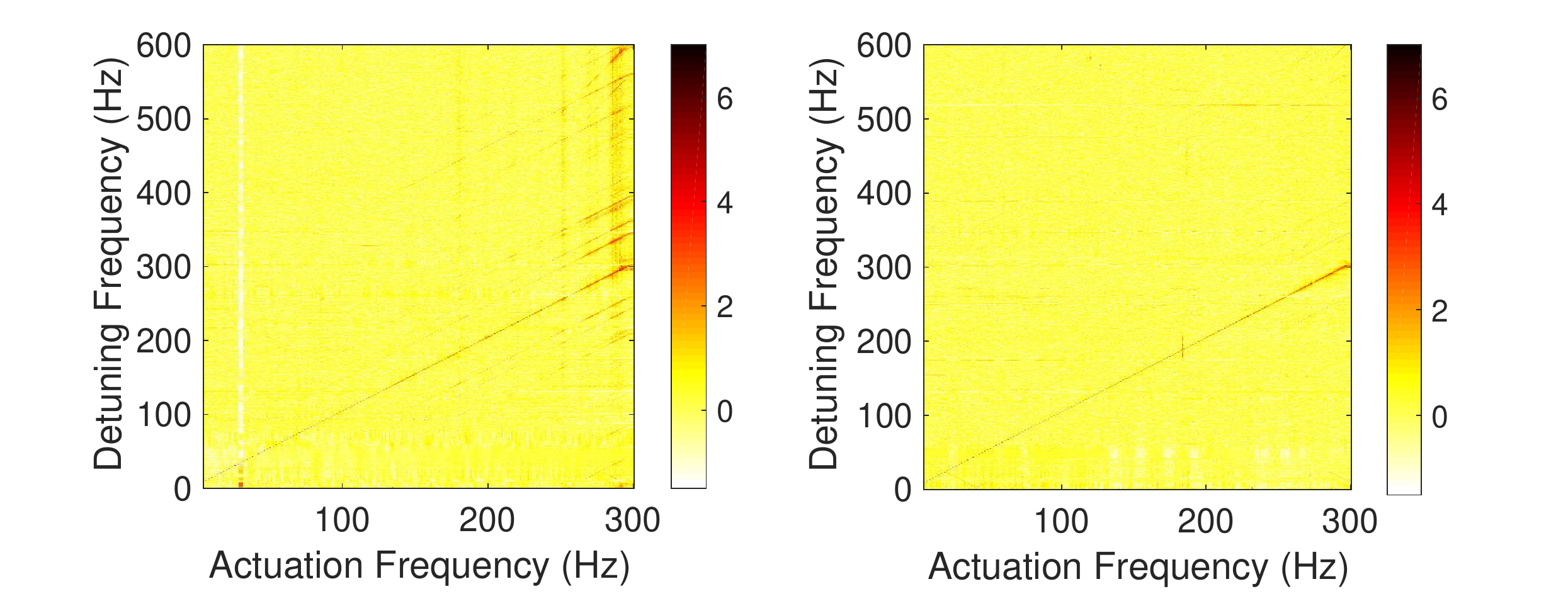}
    \caption{Spectral response of tuners to single frequency sinusoidal excitations in two cavities of the main linac used in the CBETA project. The plots show the logarithm of power spectra (in color) of tuner response as functions of actuation frequency $f_{\rm pz}$ on the x-axes and frequency of detuning $f_{\rm detuning}$ on the y-axes. The left and right panels show measurements from an unstiffened and stiffened cavity respectively. The white line on the left panel at the actuation frequency of 30 Hz indicates an absence of valid detuning data due to a RF trip during the measurement.}
    \label{fig:tunerrespnl}
\end{figure}

\subsection{Non-linear response}
The assumption of linearity is dependent on the linearity of the piezo-electric effect, the stress strain curves and damping mechanisms in the materials involved. Although the stress strain curve and the piezo-electric effect are reasonably linear in the regime of use, slight hysteresis is generally observed in resonant frequency as the applied voltage is cycled from 0 V to high voltage back to 0 V. \cite{Posen2012ipac,Pischalnikov2015,Cichalewski2015} This implies that some non-linearity is present in the system and we should verify its magnitude. A non-linear response would excite multiple frequencies even if we excite just one \cite{Elliott2015} and this provides a simple way of diagnosing the non-linear dynamics of the system by exciting it with sine waves. We measure the power spectrum of detuning for different frequencies of excitation of the tuner and subtract the contribution from the ambient microphonics present in the system to yield an approximate spectral response function. Figure~\ref{fig:tunerrespnl} illustrate some examples of spectral responses. The linear response shows up as a line with slope of 1, i.e. the frequency of excitation equals the major frequency component of detuning. However, the plot also shows evidence of higher order responses in the form of additional frequencies in the detuning spectrum.

We can use the straight lines observed in the plots to estimate the order of non-linearity present in the system and gain some insight into its source. In general, the frequencies present in non-linear responses of a dynamical system to a sinusoidal excitation can be written as,
\begin{equation}
    f_{\rm detuning} = m f_{\rm pz} + \sum_i n_i f_{\rm {vib},i}
\end{equation}
where $f_{\rm detuning}$ and $f_{\rm {vib},i}$ are the frequencies present in the tuner response and ambient microphonics, while $f_{\rm pz}$ is the frequency of excitation. $m$ and $n_i$ are integers, with $|m|$ representing the order of the non-linear term and the addition of vibration frequencies represent parametric behaviour of the tuner dependent on external microphonics. Both the cavities show lines parallel to the linear response line indicating the presence of modulation from ambient microphonics. The unstiffened cavity further shows the second harmonic with evidence of the $m=2$ line near $f_{\rm pz} = 300~\rm{Hz}$. The strength of the non-linear responses appear to be a function of frequency with excitation frequencies of above 250 Hz showing the most activity. These observations suggest that we can ignore the non-linearity as long as we excite the tuner below 250 Hz which limits the bandwidth of the compensation system.

\section{Active Noise Control}\label{sec:anc}
Microphonics compensation of SRF cavity detuning using fast tuners has been demonstrated using a variety of techniques. Resonance control of CW RF cavities typically rely on feedback of microphonics detuning. In this control topology, the detuning acts as an input to the controller which generates a signal for the piezo-electric actuator which in turn affects the net microphonics detuning thus closing the loop. The transfer function of the tuner system as discussed in the previous section plays an important role in designing the controller. The traditional method of Proportional-Integral feedback has been demonstrated in various machines \cite{Conway2010,Banerjee2017,Neumann2010} and is very effective when the phase response of the tuner is a monotonous function of frequency which is typical at lower frequencies ($\leq 10\rm{Hz}$). At higher frequencies, the tuner cavity system typically has mechanical eigen-modes which introduce steps in the phase response which may possibly lead to positive feedback and instability at even modest gains. Consequently, low pass filters are used to ensure stability of the PI loop but at the cost of reducing the bandwidth. Additional band pass filters may be used in parallel to attenuate certain frequency bands, however manually adjusting them while ensuring stability is inconvenient. In order to get past this limitation, arbitrary digital control filters can be optimized specifically to compensate for a given microphonics spectrum while taking into account the exact phase response of the tuner. This has been demonstrated on the new LCLS-II cryomodules being tested at FermiLab. \cite{EinsteinCurtis2017}

In the methods described above, the tuner transfer function and the microphonics spectrum are first measured and the data is processed external to the RF control system and the optimal filter coefficients are then uploaded into the control system. In contrast, adaptive tuning of digital control filters inside the RF system during operations using Least Mean Squares algorithms have also been demonstrated. In traditional LMS, an external reference signal which correlates to microphonics detuning is used as an input to a Finite Impulse Response (FIR) filter whose coefficients are updated continuously to reduce the mean square of detuning \cite{Neumann2010}. In a different technique \cite{Rybaniec2017} based on Active Noise Control (ANC) methods, amplitude and phase of sinewaves at different frequencies are adjusted to cancel out microphonics. However, both these methods require prior measurement of the tuner transfer function which may be a function of tuner position \cite{Posen2012PRAB} and may not stay constant over long periods of time and over multiple pressure or temperature cycles. In this paper, we derive the narrow band ANC technique and propose a modification so that it adapts to the tuner response phase in-situ.

Microphonics detuning due to narrow band vibration sources can be well approximated by a finite series of sinusoids at different frequencies with slowly changing amplitudes and phases. This motivates the use of an algorithm which works by adjusting the amplitudes of a series of sine and cosine functions in order to reduce the mean square detuning. At a particular frequency $\omega_m$, the ideal phase of the actuator signal $\theta^{\rm pz}_m$ is determined by not only the relative phase of external detuning with respect to the internal clock $\theta^{\rm micro}_m$ of the control system but also the phase response $\phi_m$ of the actuator. The ideal actuator signal phase given by $\theta^{\rm pz}_m = \theta^{\rm micro}_m + \phi_m - \pi$ in principle perfectly cancels the sine wave produced by external vibrations. The phase lag $\phi_m$ introduced by the tuner can be assumed to be a constant when the frequency of vibrations is far from a mechanical resonance, and used as a compensation parameter in the algorithm. Using the technique of stochastic gradient descent, we derive a set of equations which updates the amplitude and phase of individual sinusoids along with online optimization of the phase parameter $\phi_m$ at the frequencies of vibration.

\subsection{Derivation}
Microphonics from narrow band vibration sources may be represented by a finite series of sinusoids at different frequencies with slowly changing amplitudes and phases. Hence, in the time domain, the actuator voltage ($u_{\rm pz} (t)$) can also be written as a sum of sinusoids with frequencies $\omega_m$ and whose amplitude and phase are determined by $I_m(t)$ and $Q_m(t)$,
\begin{equation}
    u_{\rm pz} (t) = \sum_m I_m(t) \cos(\omega_m t) + Q_m(t) \sin(\omega_m t)
    \label{eq:upz}
\end{equation}
The piezo-electric actuator tunes the cavity in response to this signal, the effect of the tuner being represented as a linear transfer function $\tau (\omega)$. Using a phasor notation for the actuator voltage $u_{\rm pz} (t) \equiv \sum_m \rm{Re} \{ \tilde{A}_m(t)\rm{e}^{i \omega_m t} \}$, we can write detuning near a particular frequency as a linear response integral.
\begin{equation}
    f_m(t) = \rm{Re} \bigg\{ \frac{1}{2\pi}\int_{-\infty}^{\infty} \rm{d}\omega \int_{-\infty}^{\infty} \rm{d}t' \tilde{A}_m(t')\rm{e}^{i (\omega_m - \omega) t'} \tau (\omega) \rm{e}^{i \omega t}  \bigg \}
    \label{eq:fm}
\end{equation}
Since the spectral content of microphonics detuning is assumed to be concentrated around certain frequencies, only parts of the transfer function are relevant in modelling the tuner movements. Far from resonance, we approximate the tuner transfer function around the frequency $\omega_m$ as,
\begin{equation}
    \tau(\omega) \simeq \tau_m \rm{e}^{-i\{\phi_m + \frac{\rm{d}\phi}{\rm{d}\omega} \big\vert_{\omega_m} (\omega - \omega_m)\}}
    \label{eq:tuner}
\end{equation}
where we assumed a constant amplitude response $\tau_m$ and a phase response up to first order. Using this ansatz in Eq.~(\ref{eq:fm}) and changing the order of integration, we get,
\begin{equation}
    \begin{split}
        f_m(t) &\simeq \rm{Re} \bigg\{\int_{-\infty}^{\infty} \rm{d}t' \tilde{A}_m(t') \rm{e}^{i \omega_m  t'} \\
        & \times \frac{1}{2\pi} \int_{-\infty}^{\infty} \rm{d}\omega \, \tau_m \rm{e}^{i\{-\phi_m + \frac{\rm{d}\phi}{\rm{d}\omega} \big\vert_{\omega_m} \omega_m\}} \rm{e}^{i\omega \big(t - t' - \frac{\rm{d}\phi}{\rm{d}\omega} \big\vert_{\omega_m}\big)} \bigg\} \\
        &= \rm{Re} \bigg\{ \tau_m \rm{e}^{i\{-\phi_m + \frac{\rm{d}\phi}{\rm{d}\omega} \big\vert_{\omega_m} \omega_m\}} \int_{-\infty}^{\infty} \rm{d}t' \tilde{A}_m(t') \rm{e}^{i \omega_m  t'} \\
        & \times \delta \big(t - t' - \frac{\rm{d}\phi}{\rm{d}\omega} \big\vert_{\omega_m}\big) \bigg\} \\
    \end{split}
\end{equation}
Where the integral over $\omega$ becomes a delta function which represents the approximate time domain impulse response valid when the frequency of actuation is $\omega_m$. Using the delta function to evaluate the convolution integral, we get,
\begin{equation}
        f_m(t) \simeq \rm{Re} \{ \tau_m \tilde{A}_m(t - D_m) \rm{e}^{i(\omega_m t - \phi_m)} \}
\end{equation}
Where we have introduced the group delay $D_m \equiv \frac{\rm{d}\phi}{\rm{d}\omega} \big\vert_{\omega_m}$. The effective detuning $\delta f_{\rm comp}(t)$ of the cavity in response to the perturbation given in Eq.~(\ref{eq:upz}) is thus given by,
\begin{align}
    \begin{split}
        \delta f_{\rm comp}(t) &= \delta f_{\rm ext}(t) + \\
        & \sum_m \tau_m \{ I_m(t - D_m) \cos(\omega_m t - \phi_m) \\
        & + Q_m(t - D_m) \sin(\omega_m t - \phi_m) \}
    \end{split}
    \label{eq:model}
\end{align}
where we have combined the tuner response at different frequencies and $\delta f_{\rm ext}(t)$ is the microphonics detuning coming from external vibrations. This multi-frequency model approximates the dynamics of the tuner in the limit of narrow band oscillations where, $\frac{1}{I_m} \frac{\rm{d}I_m}{\rm{d}t}, \frac{1}{Q_m} \frac{\rm{d}Q_m}{\rm{d}t} << \omega_m$. Now we can use this model to construct a suitable cost function which can be minimized by the algorithm.

\begin{figure}
    \centering
    \includegraphics[scale=0.4]{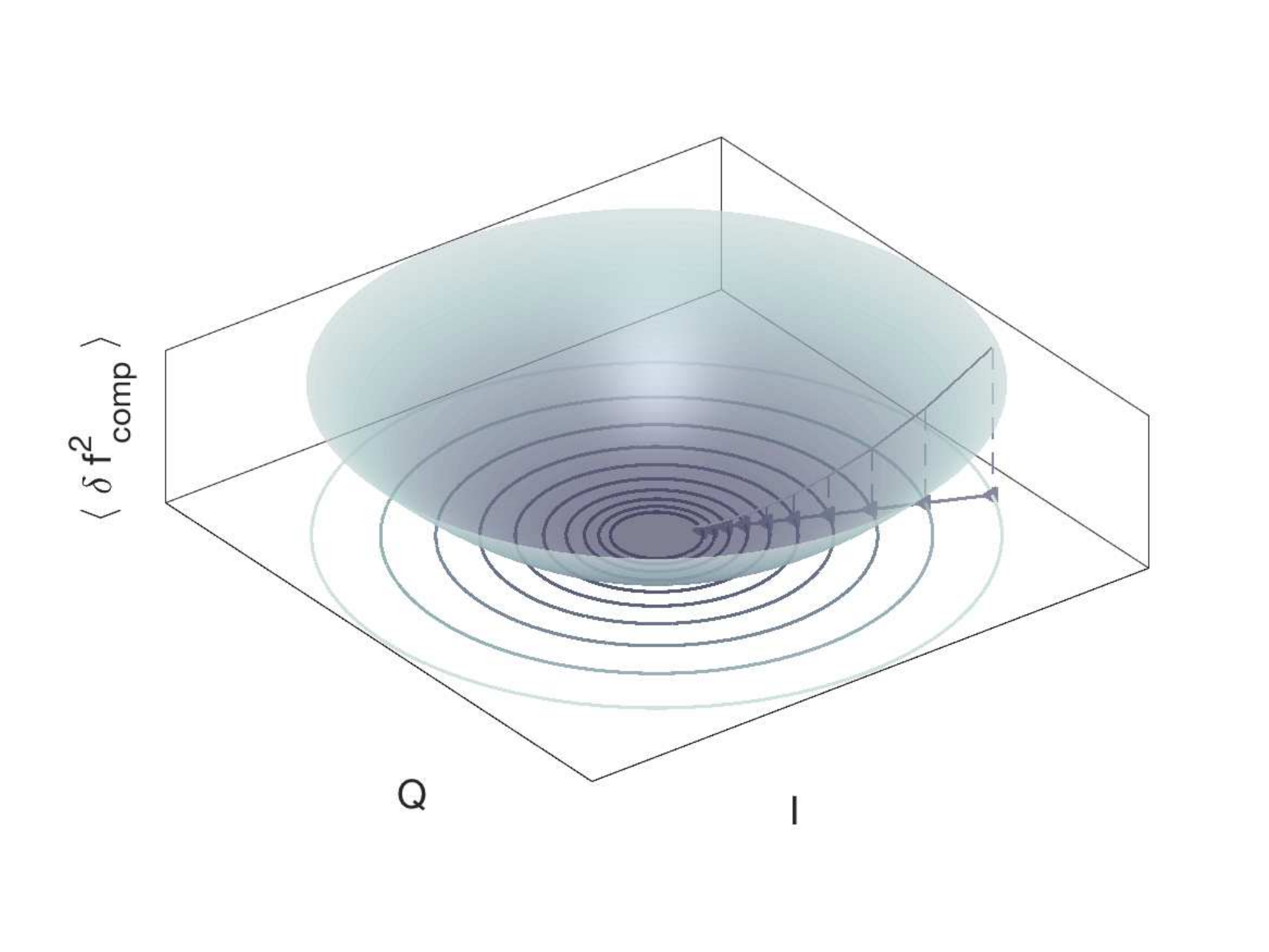}
    \caption{Demonstration of the stochastic gradient descent technique. I and Q are the control variables and $\langle \delta f^2_{\rm comp} \rangle$ is the cost function to be minimized, shown as a surface. Starting from the initial point, each successive iteration goes in the direction opposite to the gradient vector, which are normal to the equal cost contours.}
    \label{fig:lmsdemo}
\end{figure}

\begin{figure}
    \centering
    \includegraphics{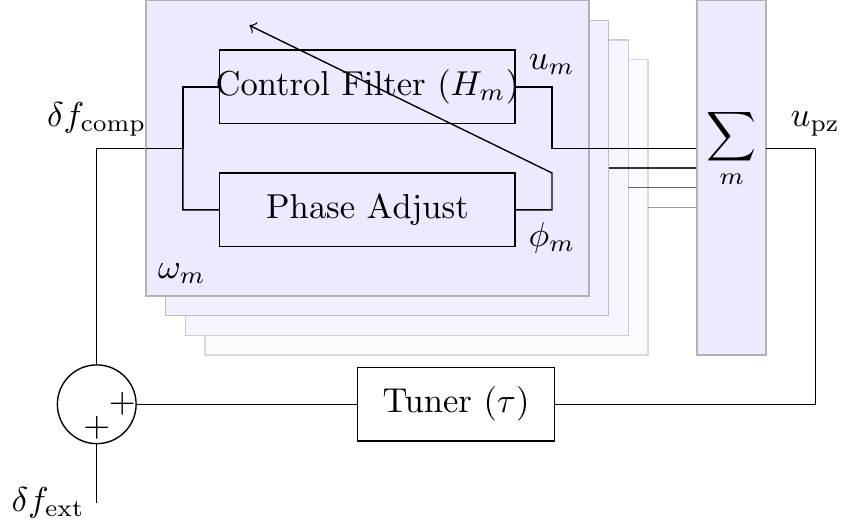}
    \caption{Schematic for the modified ANC algorithm showing the effective control filter [Eq.~(\ref{eq:Iadj}), (\ref{eq:Qadj}) and (\ref{eq:upz})] and the phase adjustment [Eq.~(\ref{eq:phiadj})] for each microphonic spectral line at $\omega_m$. The output of all the filters $u_m$ are summed and sent to the piezo-electric actuator, which tunes the cavity through a transfer function $\tau$. The compensated detuning $\delta f_{\rm comp}$ is the sum of the contributions from vibrations $\delta f_{\rm ext}$ and the tuner movements and is used as input to the control filters thus closing the feedback loop.}
    \label{fig:ControlDiag}
\end{figure}

The objective of microphonics compensation is to reduce the mean square detuning of the cavity.
\begin{equation}
    C(t_n) \equiv \frac{1}{N} \sum_{i=n - N + 1}^n (\delta f_{\rm comp}(t_i))^2
    \label{eq:meansq}
\end{equation}
$C(t_n)$ is the cost function at time $t_n$ which is taken to be the expected value of square of detuning, approximated by a running average. The method of gradient descent relies on the gradient vector being the direction of steepest descent on the cost surface as shown in Fig.~\ref{fig:lmsdemo}. In parameter space, the gradient represents the normal to the constant cost surface and can be estimated using the model developed in Eq.~(\ref{eq:model}). Following standard LMS techniques \cite{Kuo1997}, we take $N = 1$ and calculate the partial derivatives of the cost function with respect to the optimization parameters $I_m$ and $Q_m$ which determine the actuator voltage.
\begin{subequations}
\begin{align}
    \frac{\partial C}{\partial I_m}  &= 2 \tau_m \delta f_{\rm comp}(t_n)\cos(\omega_m t_n - \phi_m ) \\
    \frac{\partial C}{\partial Q_m}  &= 2 \tau_m \delta f_{\rm comp}(t_n)\sin(\omega_m t_n - \phi_m )
\end{align}
\end{subequations}
Where we have used Eq.~(\ref{eq:model}) under the assumption that group delay $D_m$ is negligible with respect to the time scales with which $I_m(t)$ and $Q_m(t)$ change. The typical group delay introduced by the modified Saclay - I tuners used in the CBETA project is less than a millisecond far from resonant frequencies while the bandwidth of the vibration sources are typically less than 1 Hz. This implies that $I_m(t)$ and $Q_m(t)$ change with time scales of more than 1 second while the delay in the feedback loop is less than a millisecond thus ensuring the validity of the above assumption. Each iteration of the stochastic gradient descent algorithm changes the control parameters a little in the direction opposite to the gradient.
\begin{subequations}
\begin{align}
    \label{eq:Iadj}
    I_m(t_{n+1}) &= I_m(t_n) - \mu_m \delta f_{\rm comp}(t_n) \times \nonumber \\
    &\cos(\omega_m t_n - \phi_m (t_n)) \\
    \label{eq:Qadj}
    Q_m(t_{n+1}) &= Q_m(t_n) - \mu_m \delta f_{\rm comp}(t_n) \times \nonumber \\
    &\sin(\omega_m t_n - \phi_m (t_n))
\end{align}
\label{eq:controller}
\end{subequations}
Where we have absorbed $2\tau_m$ into $\mu_m$, which is the adaptation rate for $I_m$ and $Q_m$ in the algorithm. Equation \ref{eq:controller} together with \ref{eq:upz} form the Active Noise Control(ANC) algorithm which is illustrated in Fig.~\ref{fig:ControlDiag}.

\subsection{Control Filter}
\begin{figure}
    \centering
    \includegraphics[scale = 0.4]{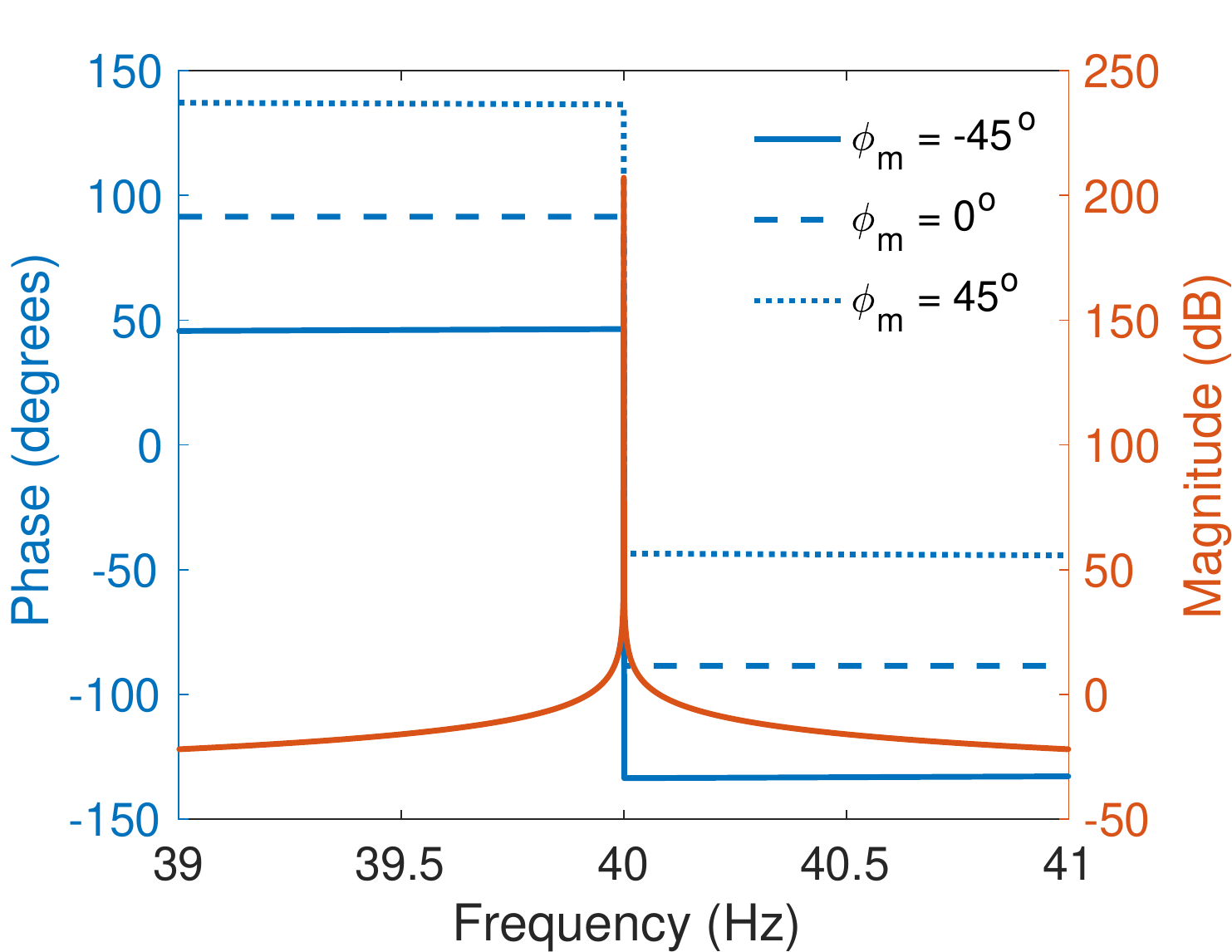}
    \caption{Frequency response of the control filter described in Eq.~(\ref{eq:LMSIR}) for $f_m = 40\,\rm{Hz}$, $\mu_m = 10^{-4}$, $\Delta t = 0.1\,\rm{ms}$ and three values of phase $\phi_m = -45^o, 0^o, 45^o$. The value of $\phi_m$ primarily provides a constant offset to the phase response and can be used as a knob to stabilize the feedback control loop.}
    \label{fig:ControlFilt}
\end{figure}

The microphonics compensation system works to attenuate multiple frequencies by effectively using an array of linear filters in parallel. Each of these control filters works on a small frequency range around $\omega_m$ and generate individual actuator signals $u_m$. $I_m$ and $Q_m$ form a complex phasor which changes slowly compared to the frequency $\omega_m$ and Eq.~(\ref{eq:upz}) modulates a sinusoid at this frequency with this phasor. Using this insight and combining the equations \ref{eq:upz}, \ref{eq:Iadj} and \ref{eq:Qadj} to eliminate $I_m$ and $Q_m$, assuming the phase $\phi_m$ does not change in time, we get
\begin{equation}
    \begin{split}
        u_m (t_{n+1}) = -\rm{Re}\{ \mu_m \rm{e}^{-i\omega_m t_{n+1}} \sum_{k=0}^n \delta f_{\rm comp}(t_k) \rm{e}^{i(\omega_m t_k - \phi_m)}\} \\
        =  - \mu_m \sum_{k=0}^n \delta f_{\rm comp}(t_{n-k}) \cos (\omega_m (k+1) \Delta t + \phi_m) \,,
    \end{split}
    \label{eq:LMSIR}
\end{equation}
where we have assumed zero initial conditions $I_m(0) = Q_m(0) = 0$. Equation~\ref{eq:LMSIR} represents a discrete convolution of the input signal with a sinusoid and is an impulse response filter. The Z-transform of the filter can be written as,
\begin{equation}
    \begin{split}
        H_{m}(z) &= \mu_m \sum_{k=0}^\infty z^{-k} \cos (\omega_m (k+1) \Delta t + \phi_m) \\
        &= \mu_m \frac{ \cos(\omega_m \Delta t + \phi_m) -  z^{-1}\cos\phi_m}{1 - 2\cos(\omega_m\Delta t)z^{-1} + z^{-2}} \,,
    \end{split}
    \label{eq:Hm}
\end{equation}
where we have dropped the minus sign for the sake of convention and $\Delta t$ is the sample duration. The frequency response of the filter is shown in Fig.~\ref{fig:ControlFilt}.

The control filter shows a very narrow band pass response rising to $\infty$ and a phase swing of $180^o$ around the frequency $\omega_m$. In the limit of $\omega \rightarrow \omega_m$, the filter response can be approximated by
\begin{equation}
    H_m(\omega) =  \frac{\mu_m\rm{e}^{i(-\frac{\pi}{2} + \phi_m + \omega_m\Delta t)}}{2(\omega - \omega_m)\Delta t}
    \label{eq:LMSIRApprox}
\end{equation}
$\mu_m$ serves as an overall scaling factor governing the span of frequencies within which the controller-tuner system has more than unity gain and $\mu_m$ can hence be used to adjust the bandwidth of the feedback loop around the microphonics frequency. In the neighborhood of $\omega_m$, the phase swings from $\pi/2 + \phi_m + \omega_m\Delta t$ to $-\pi/2 + \phi_m + \omega_m\Delta t$ from left to right in a discrete jump. This span of phase can be adjusted to stabilize the controller and optimize compensation.

\begin{figure}
    \centering
    \includegraphics[scale=0.4]{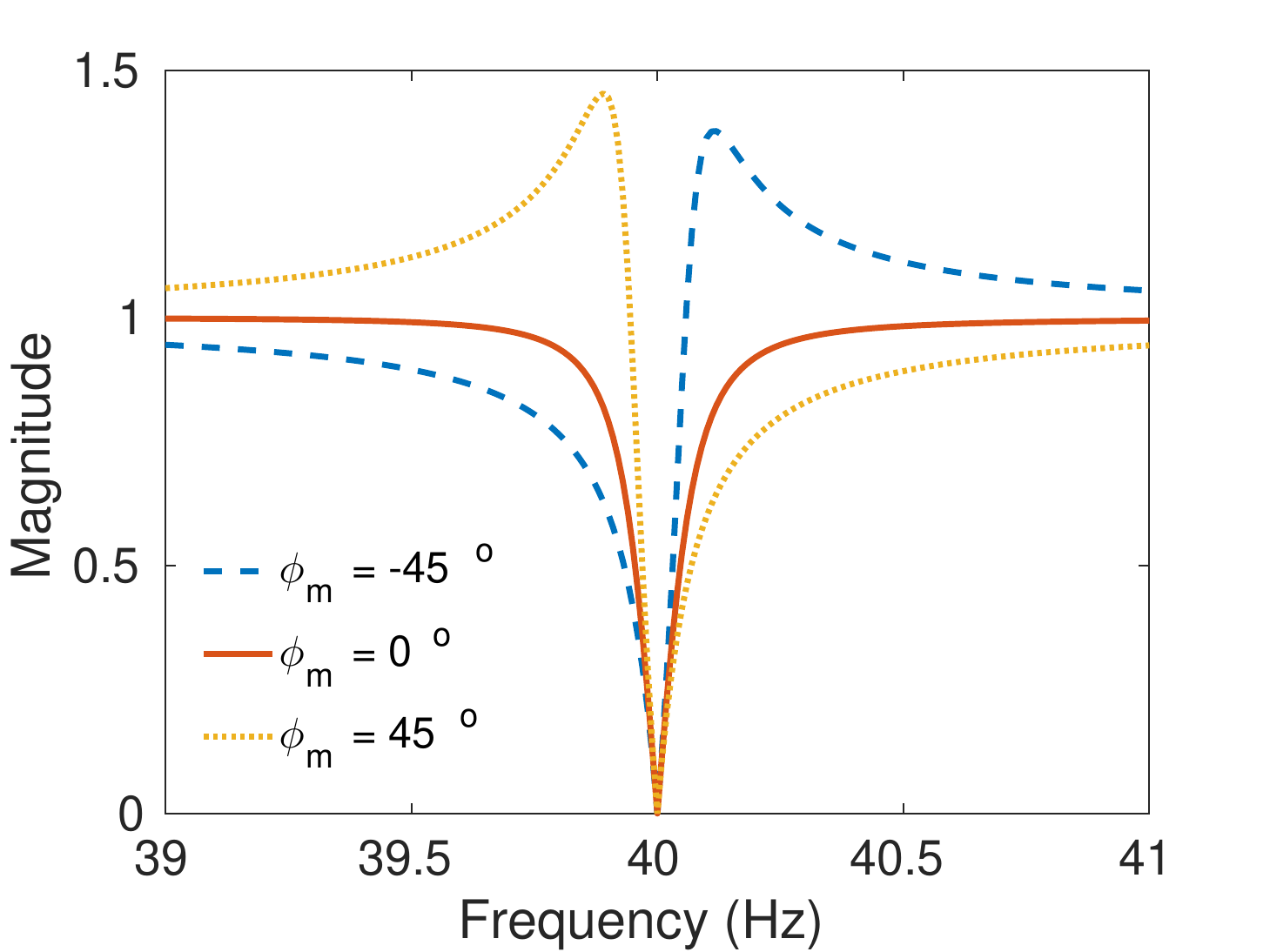}
    \caption{Magnitude of closed loop transfer function in frequency domain with $f_m = 40\,\rm{Hz}$, $\mu_m = 10^{-4}$, $\Delta t = 0.1\,\rm{ms}$ for different choices of $\phi_m$ when $\tau(\omega) = 1$. Regardless of the value of $\phi_m$, microphonics at the frequency $f_m$ is perfectly compensated.}
    \label{fig:CLTF}
\end{figure}

The compensated detuning $\delta \tilde{f}_{\rm comp}(\omega)$ is the net effect of the tuner $\delta \tilde{f}_m(\omega) \equiv \tau(\omega) u (\omega)$ and the external contribution $\delta \tilde{f}_{\rm ext}(\omega)$. The tuner excitation $u(\omega) \equiv -H(\omega)\delta \tilde{f}_{\rm comp}(\omega)$ is obtained as an output of the linear controller whose frequency response is given by $H(\omega)$, where the minus sign accounts for the one we dropped while deriving Eq.~(\ref{eq:Hm}). From these definitions we obtain the closed loop transfer function (CLTF) of the system which provides a linear relation between the external detuning $\delta \tilde{f}_{\rm ext}(\omega)$ and the compensated detuning $\delta \tilde{f}_{\rm comp}(\omega)$ in frequency space.
\begin{equation}
        \mathcal{C}(\omega) \equiv \frac{\delta \tilde{f}_{\rm comp}(\omega)}{\delta \tilde{f}_{\rm ext}(\omega)} = \frac{1}{1 + \sum_m H_m(\omega) \tau(\omega)}
        \label{eq:cltf}
\end{equation}
Where the sum of all the filters $H(\omega) \equiv \sum_m H_m (\omega)$ acts as a comb with it's frequency response amplitude remaining small except for the neighborhood of $\omega_m$. In a theoretical situation, when all the microphonics detuning is generated by pure sine waves whose frequencies are exactly $\omega_m$, the controller works perfectly to compensate for all microphonics as shown in Fig.~\ref{fig:CLTF} since $\lim_{\omega \rightarrow \omega_m} H_m(\omega) = \infty$. However real microphonics signals have finite bandwidth spectral modes and the performance of feedback control is determined by the combined response of the controller and the tuner over all of frequency space. In the limit of the response of each filter being much greater near its pass-band than its nearest neighbours i.e. $|\tau(\omega_m)H_{m-1}(\omega_m)| << 1$, we can approximate the net transfer function to be a product of individual response functions at each compensation frequency.
\begin{equation}
        \mathcal{C}(\omega) \sim \prod_m \frac{1}{1 + H_m(\omega) \tau_m(\omega)}
        \label{eq:cltfprod}
\end{equation}
Figure \ref{fig:CLTF} shows the shape of one such isolated function and illustrates how adjusting $\phi_m$ can lead to an asymmetric response, attenuating vibrations on one side and amplifying the other. Further, the closed loop stability of the individual contributions with index $m$ can be a sufficient condition for the stability of the entire system. 

The region of stability of the feedback loop around each frequency $\omega_m$ determines the range of acceptable values of $\mu_m$ and $\phi_m$. Using the tuner response [Eq.~(\ref{eq:tuner})] and the approximate transfer function of the filter [Eq.~(\ref{eq:LMSIRApprox})], the phase $\phi_{\rm{OL}}$ of the open loop transfer function $\mathcal{U}_m(\omega) \equiv H_m(\omega)\tau(\omega)$ around the frequency $\omega_m$ is given by,
\begin{equation}
    \phi_{\rm{OL}} = 
    \begin{cases}
        \frac{\pi}{2} + \phi_m + \omega_m\Delta t - \phi'_m & - (\omega - \omega_m) D_m, \\
        & \text{for } \omega \leq \omega_m \\
        -\frac{\pi}{2} + \phi_m + \omega_m\Delta t - \phi'_m  & - (\omega - \omega_m) D_m, \\
        & \text{for } \omega > \omega_m \\
    \end{cases}
\end{equation}
Where we have used $\phi'_m$ and $D_m$ as the transfer function phase and group delay at frequency $\omega_m$. $\phi_m$ is the compensation parameter used in the control algorithm. The band of frequencies for which $|\mathcal{U}_m(\omega)| \geq 1$ is given by,
\begin{equation}
    \omega_m - \frac{\mu_m \tau_m}{2\Delta t} < \omega < \omega_m + \frac{\mu_m \tau_m}{2\Delta t}
    \label{eq:ugain}
\end{equation}
The control loop is stable when we avoid positive feedback inside the above domain i.e. $\phi_{\rm{OL}}(\omega_m - 0.5\mu_m\tau_m/\Delta t) < \pi$ and $\phi_{\rm{OL}}(\omega_m + 0.5\mu_m\tau_m/\Delta t) > -\pi$. This gives us a range of possible values for $\phi_m$.
\begin{equation}
    \begin{split}
        - \frac{\pi}{2} + \frac{\mu_m\tau_m D_m}{2\Delta t} - \omega_m\Delta t + \phi'_m < \phi_m \\
        < \frac{\pi}{2} - \frac{\mu_m\tau_m D_m}{2\Delta t} - \omega_m\Delta t + \phi'_m
    \end{split}
    \label{eq:PhiSpan}
\end{equation}
The center of the above range, $\phi_m^{\rm center} =  \phi'_m - \omega_m\Delta t$ compensates for the phase lag from the tuner at frequency $\omega_m$ and gives us the maximum margin on stability. The span of acceptable values of $\phi_m$ depends on not only the tuner behavior but also the adaptation rate $\mu_m$. To ensure that the range given by Eq.~( \ref{eq:PhiSpan}) is not a null set, we put an upper bound on $\mu_m$.
\begin{equation}
    \mu_m < \frac{\pi \Delta t}{D_m \tau_m}
    \label{eq:MuSpan}
\end{equation}
These calculations assume that neighbouring frequencies of the comb are far away so that their response amplitudes are much lesser than unity at the next frequency i.e. $|\tau(\omega_m)H_{m-1}(\omega_m)| << 1$. This gives us a crude limit on the distance between nearest neighbours as,
\begin{equation}
     |\omega_m - \omega_{m-1}| >> \frac{\mu_{m-1}\tau_m}{2\Delta t}~,
\end{equation}
hence constraining the spacing of the different frequencies we can compensate. Detailed calculations involving the complete tuner transfer function and the array of band pass filters are required to fully analyze the stability of the control system.

\begin{figure}
    \centering
    \includegraphics[scale=0.4]{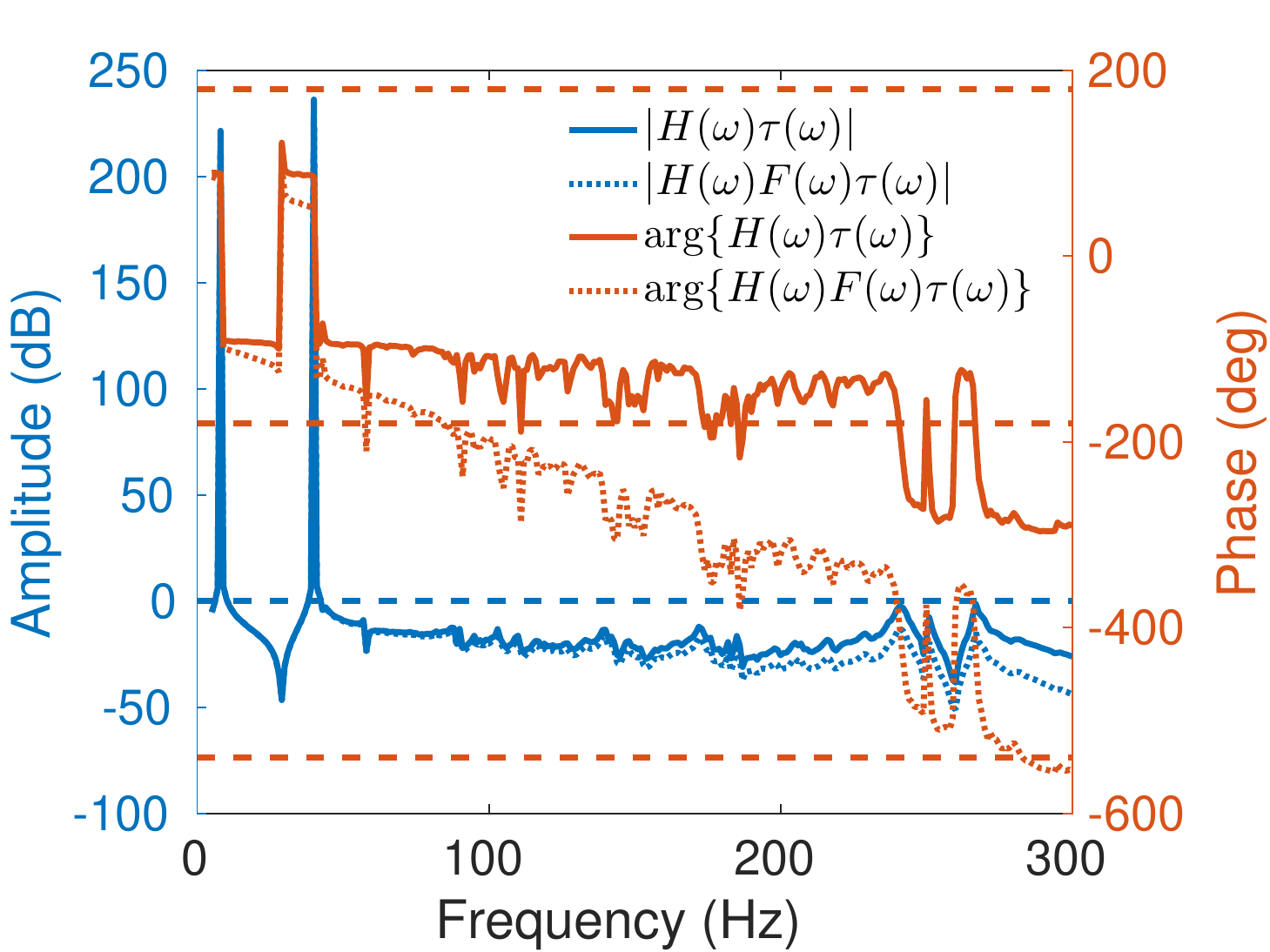}
    \caption{Bode plot for the mechanical open loop transfer function $\mathcal{U}(\omega) \equiv H(\omega)\tau(\omega)$ of an unstiffened cavity inside the main linac used for the CBETA project, showing both the amplitude and phase in blue and orange respectively. The solid lines represent the effect of the Active Noise Control algorithm applied to frequencies 8 Hz and 40 Hz, while the dotted lines represent the effect of incorporating a low pass Finite Impulse Response filter with frequency response $F(\omega)$ inside the controller. The blue dashed line represents unity gain (0 dB) and the orange dashed lines at $-540^o, -180^o \text{ and } 180^o$ represents the boundaries of stability in phase.}
    \label{fig:ancstab}
\end{figure}

\begin{figure*}[t]
    \centering
    \includegraphics[scale=0.5]{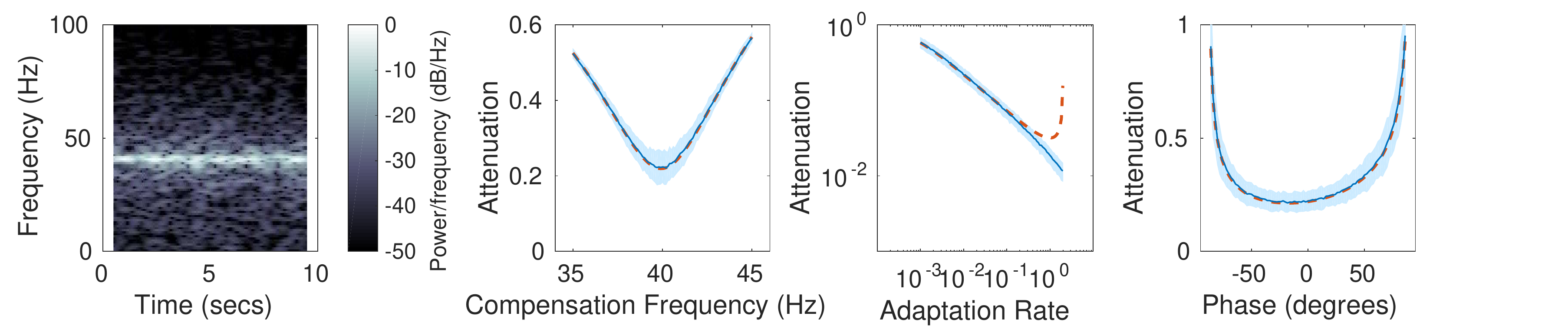}
    \caption{Simulation results of using ANC with an ideal tuner. From the left, the first panel shows the spectrogram of simulated vibrations and the others show dependence of $\sqrt{\langle \delta f^2_{\rm comp}\rangle_t /  \langle \delta f^2_{\rm ext}\rangle_t}$  on frequency $\omega_m$, adaptation rate $\mu_m$ and controller phase $\phi_m$ respectively. The average of attenuation over an ensemble of 100 simulations is shown as the thin blue line and its $2\sigma$ confidence bounds are shaded light blue. The dashed red lines show the prediction from the model.}
    \label{fig:paramSim}
\end{figure*}

The complete stability analysis of the ANC system involves assessing the open loop transfer function over all frequencies using a bode plot. Figure \ref{fig:ancstab} shows an example of using the compensation system on an unstiffened cavity used in the main linac of the CBETA project. We apply the algorithm on two frequencies 8 Hz and 40 Hz illustrated by the notches in amplitude and expected phase swings of $180^o$ at these frequencies. The open loop phase stays between the $-180^o$ and $180^o$ lines for gains above 0 dB showing that the system is stable near these frequencies. The phase margins i.e the distances from the $-180^o$ line when the amplitude crosses unity gain (0 dB) are $80^o$ and $90^o$ at 8 Hz and 40 Hz respectively as seen from the plot of $\phi_{\rm{OL}} = \rm{arg}\{H(\omega)\tau(\omega)\}$ in Fig.~\ref{fig:ancstab}. However, the gain seems to be close to 0 dB near the tuner resonances at frequencies around 250~Hz, when $\phi_{\rm OL}$ crosses the $-180^o$ mark with a gain margin $\lesssim 2$ dB. This prompts the use of a low pass filter with frequency response $F(\omega)$ to attenuate the transfer function at these frequencies as shown by the dotted lines of Fig.~\ref{fig:ancstab}. The analysis illustrates the effect of tuner resonances far from the compensation frequencies $\omega_m$ signalling the need for additional filtering to ensure stability of the system.

The performance of the controller in suppressing microphonics detuning depends on the vibrating components. Assuming that the microphonics is generated by resonant processes either through instabilities or white noise excitations, the ensemble averaged power spectrum $\langle |\delta \tilde{f}_{\rm ext} (\omega)|^2 \rangle_{\rm{E}}$ is given by,
\begin{equation}
    \langle |\delta\tilde{f}_{\rm ext} (\omega)|^2 \rangle_{\rm{E}} \equiv \sum_v \frac{\Gamma_v^2}{\big \{1 - (\frac{\omega}{\omega_v})^2 \big \}^2 + \big (\frac{\omega}{Q_v \omega_v} \big )^2} \, ,
\end{equation}
where $\omega_v$, $Q_v$ and $\Gamma_v$ are the frequencies, quality factors and strengths of microphonics detuning. Figure \ref{fig:paramSim} shows the effect of adjusting compensation parameters from simulations of the controller assuming a perfect tuner with $\tau(\omega) = 1$, group delay $D = 2\Delta t$ and random microphonics detuning with frequency 40~Hz and quality factor 50 as illustrated by the first panel. To account for the randomness, we perform the ensemble average of attenuation over 100 simulations each lasting for a duration of 10~seconds. We calculate the expected performance of the algorithm using the closed loop transfer function [Eq.~(\ref{eq:cltfprod})] along with the filter response in Eq.~(\ref{eq:Hm}). The results show that the controller performs its best on average when $\omega_m = \omega_v$, with attenuation progressively getting worse as we go farther away from $\omega_v$. The attenuation shows an asymmetric dependence on $\phi_m$ about $0^o$ reaching a minimum at some non-zero value. This leads to a closed loop transfer function asymmetric about $\omega_v$ which exactly corrects for the asymmetry of the vibration power spectrum. Finally, $\mu_m$ represents the gain in the system and compensation is expected to get better with larger gain up to the limit given by Eq.~(\ref{eq:MuSpan}) beyond which the system becomes unstable. The expected attenuation clearly diverges at $\mu_m \gtrsim \pi/2$, however the results from the numerical simulations don't agree. In practice the maximum gain of the system will depend on the exact response of the tuner, especially the group delay. These simulations guide us on how to choose parameters of the ANC during operations.

\subsection{Phase Adaptation}
The compensation performance of a controller with fixed parameters is dependent on variations in the response of the tuner and fluctuations of the microphonics spectrum. The tuner response may vary from day to day due to pressure variation in the Helium bath while the vibration mechanism may also change frequency as a function of time. The controller as described in the previous section will not be able to adapt to such changes, which might limit performance in a dynamic environment. The simulation results in Fig.~\ref{fig:paramSim} suggests that attenuation is a monotonously decreasing function of gain $\mu_m$, with the system becoming unstable beyond a threshold. An adaptive algorithm to optimize for the value of $\mu_m$ might tend to drive the system towards instability. However, the controller frequency $\omega_m$ and phase $\phi_m$ have positions where attenuation is minimum within the range of values which satisfy the stability conditions. Consequently, adapting $\omega_m$ and $\phi_m$ to a changing excitation could potentially make the algorithm more robust, while making it easier to operate in practice since it would optimize itself.

\begin{figure}[h]
    \centering
    \includegraphics[scale=0.5]{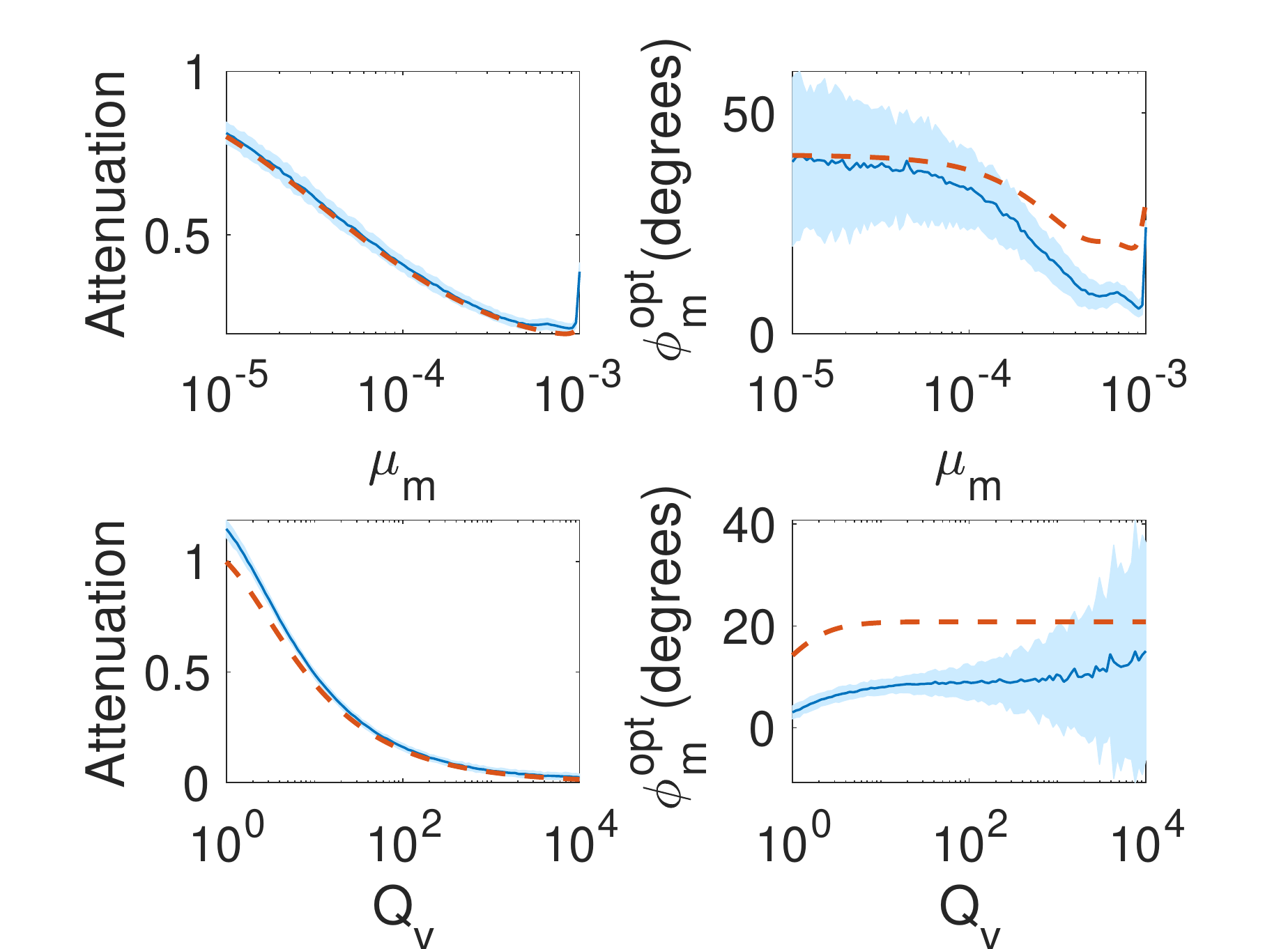}
    \caption{Comparison of controller phase optimization for different gains and quality factors of vibration. (a) and (c) show the best attenuation reached while (b) and (d) shows the optimum phase as functions of $\mu_m$ and $Q_v$ respectively. Results from offline optimization are shown as dashed lines while simulations of stochastic gradient descent are shown as the thin blue lines with $2\sigma$ confidence bounds shaded light blue.}
    \label{fig:phistudy}
\end{figure}

The optimization of the ANC system translates to finding the minimum of the mean square of detuning $\langle \delta f^2_{\rm{comp}} \rangle$ with respect to the phases $\phi_m$ and the frequencies $\omega_m$. For finding the optimum values, we can use the Parseval's theorem to write the mean square in frequency space and then use Eq.~(\ref{eq:cltf}) to establish a relation with $H_m(\omega)$.
\begin{equation}
    \begin{split}
        \langle \delta f^2_{\rm{comp}} \rangle_t &\propto \int_{-\infty}^{\infty} |\delta \tilde{f}_{\rm comp} (\omega)|^2 \rm{d}\omega \\
        &= \int_{-\infty}^{\infty} \bigg|\frac{\delta \tilde{f}_{\rm ext} (\omega)}{1 + \sum_m H_m(\omega)\tau(\omega)}\bigg|^2 \rm{d}\omega
    \end{split}
    \label{eq:mscompfreq}
\end{equation}
We can use this expression with any numerical optimizer to calculate the best values for the ANC parameters. The dashed lines in Fig.~\ref{fig:phistudy} show the optimum values of phase $\phi_m^{\rm opt}$ and the attenuation calculated using a simplex algorithm assuming microphonics at 40 Hz. Further, the calculations also use a tuner transfer function fitted to data from MLC un-stiffened cavity 3. Consequently any LMS scheme based on this approach will adapt to changing characteristics of microphonics detuning but not to changes in tuner response. We can derive an alternative algorithm which works in the time domain and does not need prior measurements of the tuner transfer function.

The cost function may be minimized with respect to controller phase $\phi_m$ using stochastic gradient descent. The partial derivative of the cost function given in Eq.~(\ref{eq:meansq}) with respect to $\phi_m$ is given by,
\begin{equation}
    \begin{split}
        \frac{\partial C}{\partial \phi_m } = 2\tau_m\delta f_{\rm comp}(t_n) \times \\
        \{I_m(t_n)\sin(\omega_m t_n - \phi_m(t_n)) \\
         - Q_m(t_n)\cos(\omega_m t_n - \phi_m(t_n))\} \, .
    \end{split}
\end{equation}
This gives us the update rule,
\begin{equation}
    \begin{split}
            \phi_m(t_{n+1}) = \phi_m(t_n) - \eta_m\delta f_{\rm comp}(t_n) \times \\
        \{I_m(t_n)\sin(\omega_m t_n - \phi_m(t_n)) \\
         - Q_m(t_n)\cos(\omega_m t_n - \phi_m(t_n))\}
         \label{eq:phiadj}
    \end{split}
\end{equation}
where $\eta_m$ is the adaptation rate for $\phi_m$. Figure~\ref{fig:phistudy} shows the results from simulations of this algorithm with microphonics at frequency 40~Hz, with a tuner response modelled on MLC cavity 3. The thin blue lines representing the ensemble average over 80 simulations clearly show the LMS adapted phase deviates from the optimization outlined previously. This deviation arises from the approximate model [Eq.~(\ref{eq:model})] of compensated detuning which we used to construct the partial derivative. However, the attenuation obtained from gradient descent closely matches the ideal result, thus demonstrating the efficacy of this method.

\section{Results}
The Cornell-BNL ERL Test Accelerator (CBETA)\cite{Hoffstaetter2017,Hoffstaetter2017Linac,Trbojevic2017Ipac} project will be the first high-current multi-turn ERL employing SRF Linacs. It uses two SRF cryomodules, one for the injection system and the other used to execute energy recovery. The injector cryomodule\cite{Hoffstaetter2012,Hartill2011} consists of five 2-cell SRF cavities\cite{Liepe2010} and is configured to provide 6~MeV of energy gain to the electron beam for injection into the CBETA loop and is operated with a low external quality factor due to high beam loading. The main linac\cite{Eichhorn2014,Eichhorn2015Cryo} on the other hand incorporates six 7-cell SRF cavities\cite{Furuta2016} with a design energy gain of 36~MeV and will be used to execute energy recovery. Operated at $Q_L \approx 6 \times 10^7$ with solid state amplifiers, the peak detuning which can be tolerated by the main linac cavities is limited to 54~Hz with a 5~kW RF source, consequently microphonics detuning presents a significant operational bottleneck and needs to be mitigated.

\subsection{Passive Suppression}

\begin{figure*}
    \centering
    \begin{tabular}{c}
         \includegraphics[scale = 0.5]{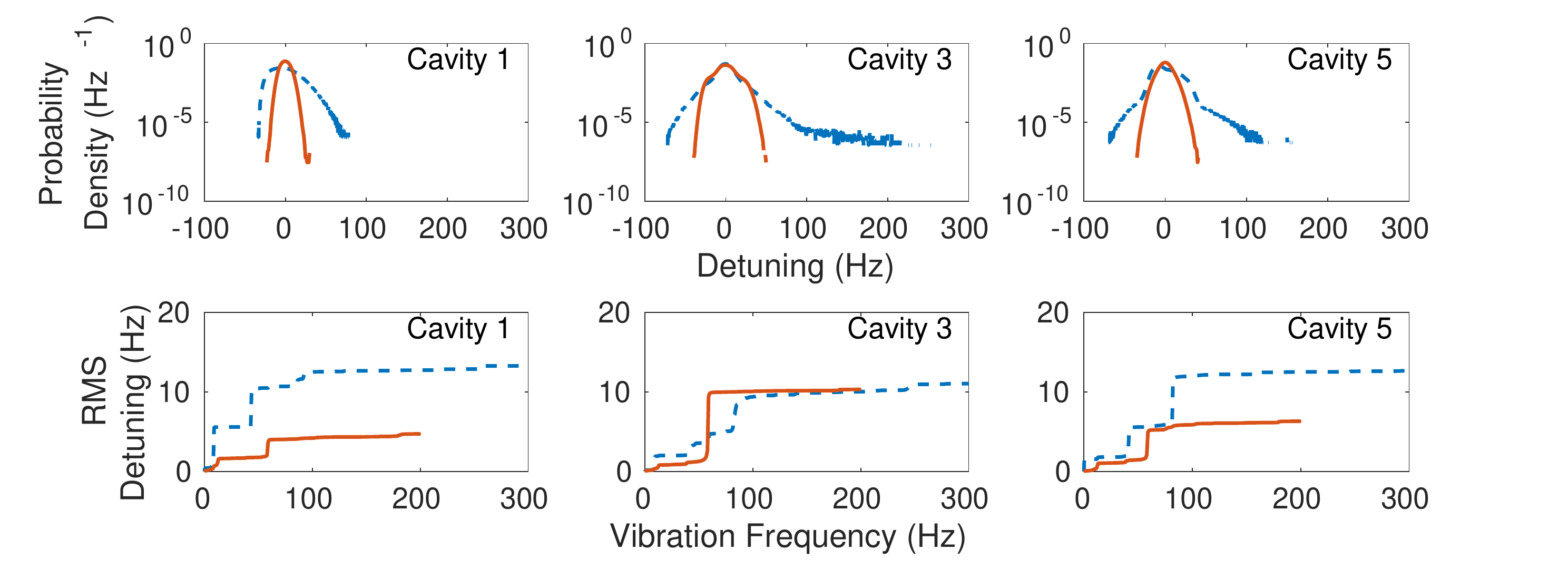}  \\
         \includegraphics[scale = 0.5]{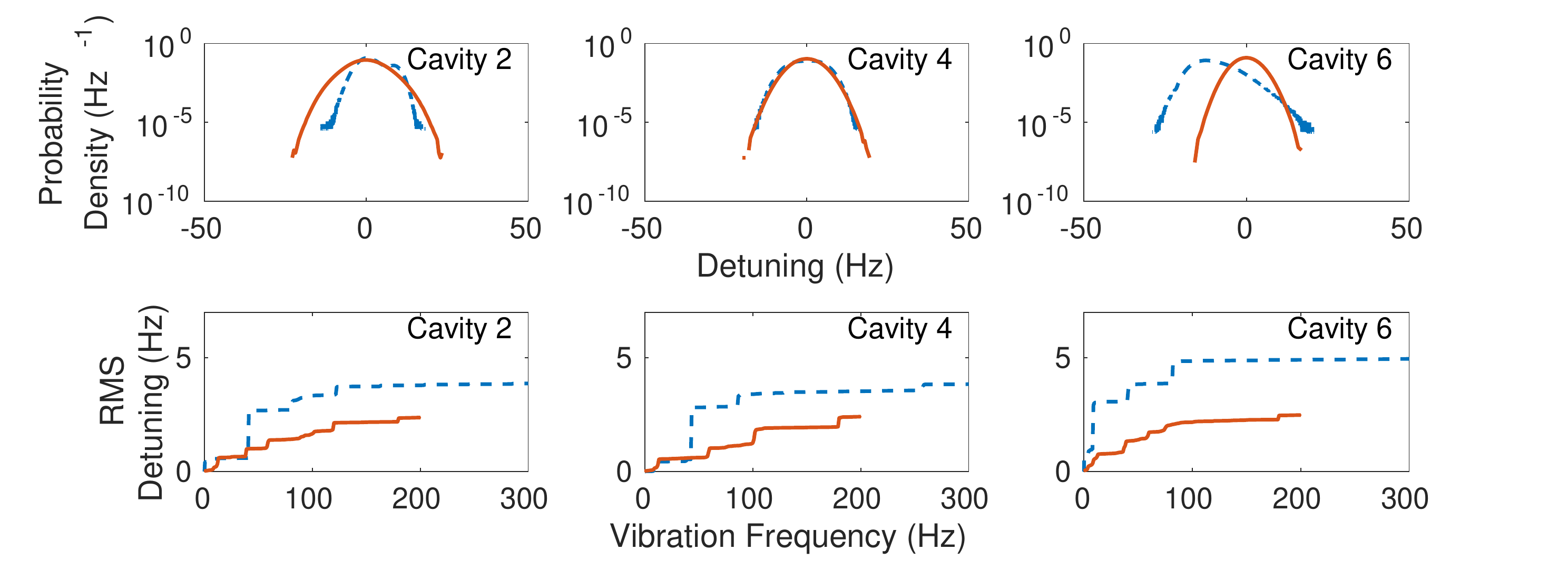}
    \end{tabular}
    \caption{Microphonics measurements on all cavities of the main linac before and after the modifications of the cryogenic system. The dashed lines represent data from the default configuration for cavities 2, 3, 5 and 6; while the data for cavities 1 and 4 were taken after making the JT and precool valves static. The solid lines indicate data after the JT and precool valves were made static and the 5~K adjust valve was fitted with sleeves.}
    \label{fig:microimprov}
\end{figure*}

The initial microphonics measurements of the main linac cavities showed strong vibrations at frequencies 8~Hz, 41~Hz and 82~Hz as illustrated in the plots of RMS detuning in Fig.~\ref{fig:microimprov}. Apart from steady vibrations at these frequencies, sudden events resulting in large peak detuning of over 100~Hz were seen in the un-stiffened cavities 3 and 5 as evident from the histograms. Vibrations can mechanically couple into the cavities from sources both inside and outside the cryomodule. In an attempt to find them, we cross-correlated the microphonics detuning signal with vibration signatures from various machinery. We started with the rotary and turbomolecular pumps maintaining the insulation vacuum in the cryomodule, looking at the effect of power cycling them for brief periods and eventually calculating the cross-correlation functions. Though the rotary pump didn't have any effect, the turbomolecular pump does induce weak vibrations around 820~Hz, consistent with a rotation speed of 50000~rpm. There are large variable frequency induction motor water pumps on the experimental floor in the vicinity of the cryomodule, these were also shown to be of no effect to the microphonics detuning. Further we also measured vibrations from the large room temperature vacuum pumps controlling the vapor pressure of Helium inside the cryomodule, showing that these too don't contribute to peak detuning of the cavities directly. Besides direct mechanical coupling of vibrations through the cavity supports, pressure fluctuations in the liquid helium surrounding the cavity also give rise to microphonics detuning. These pressure variations accounted for most of the microphonics in the main linac cavities.

The cryogenic system of the main linac cryomodule is a modified version of the TESLA design.\cite{He2012} Separate vessels house the six cavities and are supplied with liquid Helium through chimneys by the 2K - 2 phase pipe and through the precool line connected to the bottom of the vessels. The pressure exerted by liquid Helium on the cavity walls influences the resonant frequency of the cavities and needs to be regulated. Slow trends in this pressure give rise to very low frequency microphonics detuning ($\lesssim 1\rm Hz$) and tight pressure regulation requires the interplay of two control mechanisms. A Joule-Thomson (JT) valve maintains the liquid level in the 2K - 2 phase pipe and an external pump maintains the vapor pressure near 12.5~Torr corresponding to 1.8 K. Two separate proportional integral feedback loops actuate the JT valve and control the pump to maintain the liquid level and vapor pressure at their set points respectively. The system also opens the precool valve when the liquid level goes below a threshold. Consequently, transients or instabilities in any of these components may give rise to vibrations in the cryomodule.

\begin{figure}
    \centering
    \includegraphics[scale=0.32]{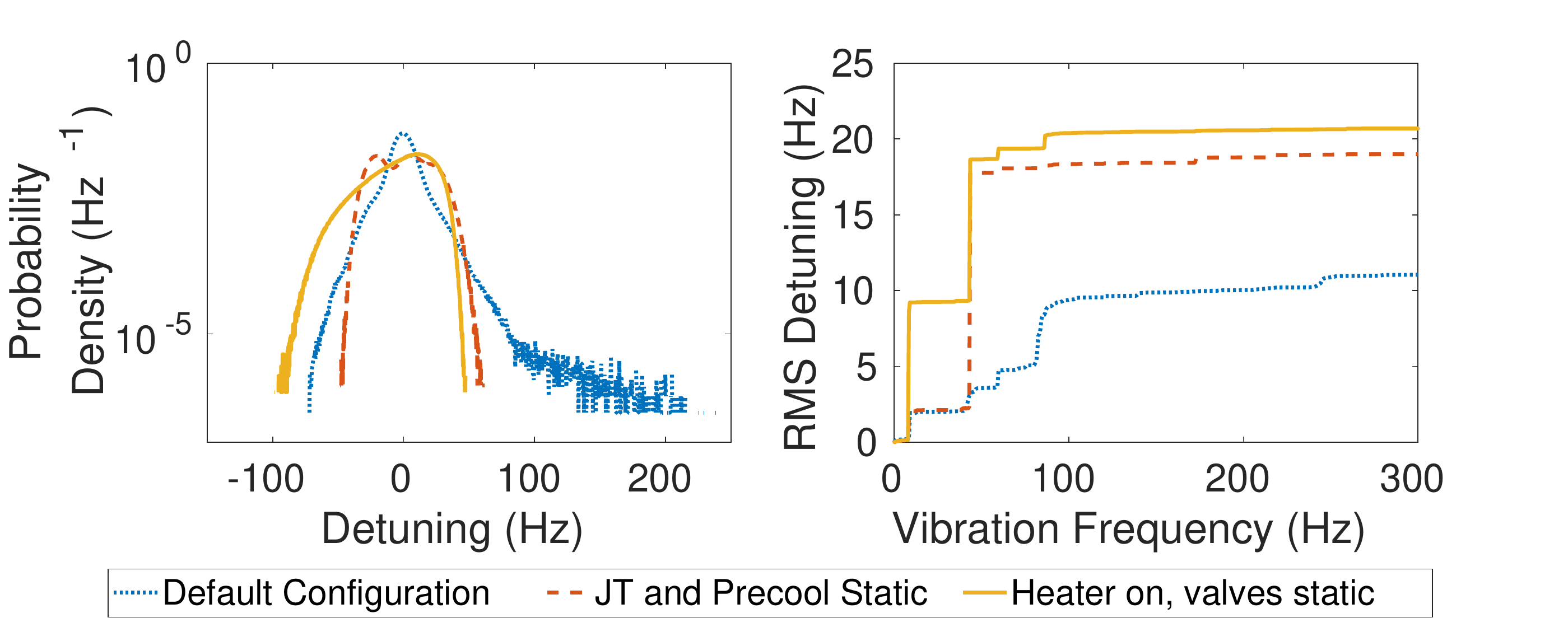}
    \caption{Influence of valve actuation on microphonics detuning of cavity 3 (unstiffened) showing the detuning histogram on the left panel and the RMS detuning on the right from measurements of duration 800 seconds.}
    \label{fig:jtexp}
\end{figure}

Measurements of microphonics detuning and various cryogenic control parameters showed that movement of the JT and movement of the precool valve both coincided with the large peak detuning events. The occasional actuation of the precool valve in response to the Helium liquid level going below a threshold correlated with spikes in a signal from a piezo-electric sensor. However the occurrence of peak detuning is more frequent, consequently we made both the precool and the JT valves static and the results of this test are shown in Fig.~\ref{fig:jtexp}. In the default configuration, the microphonics histogram shows large peak events which are $\gtrsim 200 \rm Hz$\footnote{A slight offset in the phase calibration of the field probe signal might have led to this uncharacteristically large value of peak detuning. However all data shown in Fig.~\ref{fig:jtexp} were taken with the same calibration settings and so this is a valid verification of the valve actuation effect.}, while the peak microphonics detuning becomes $\sim 50 \rm Hz$ with both valves static verifying the proposed mechanism. However, this configuration doesn't allow us to have active control on the liquid Helium level and if the boil off generated due to the thermal load from the cavities doesn't equal the rate of in-flow from the supply line, then the liquid level in the 2K - 2 phase pipe would steadily run away. To avoid this, a heater attached to the 2K - 2 phase pipe is put on a control loop to provide a minimum dynamic thermal load to substitute for when the cavities are not generating enough heat and thus boiling off suitable amounts of Helium while stabilizing the liquid level. This results in stable operation while limiting the peak detuning to $\sim 100\,\rm Hz$ a definite improvement from the original configuration. Despite the additional detuning introduced due to the operation of the heater, this is still better than controlling the Helium level using the valves.

Peak detuning was greatly improved when the valves were made static, however we observed a strong enhancement of the steady state oscillations at 41~Hz which don't contribute much to the peak detuning but increase the RMS by a factor of 2. The liquid Helium level control using the heater enhances the narrow band 8~Hz vibration line. This points to gas flow in the Helium Gas return Pipe (HGRP) as a possible source for the generation of the 8~Hz vibrations. Previous operations data further corroborated this fact by showing that the vibration amplitude at 8~Hz is an increasing function of the vapor flow through the Helium Gas Return Pipe possibly exciting a mechanical eigen-mode of the structure. Pending further investigation into the source, the active compensation system has been successful in attenuating these vibrations.

\begin{figure}
    \centering
    \includegraphics[scale = 0.8]{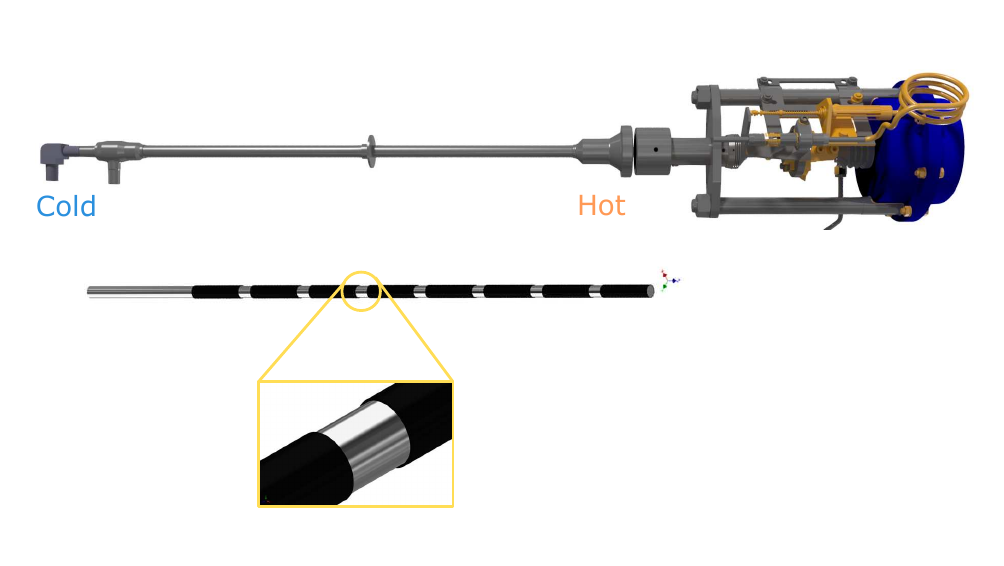}
    \caption{Cryogenic needle valve used to regulate Helium flow in the cryomodule and the electro-pneumatically actuated valve stem showing attached plastic sleeves filling the space between the stem and the inner surface of the stalk.}
    \label{fig:valveandsleeves}
\end{figure}

Accelerometer measurements of vibration on the 5~K adjust cryogenic valve stalk yielded significant cross-correlation with the microphonics detuning measurement at 41 Hz and 82 Hz. Figure \ref{fig:valveandsleeves} is a schematic of the valve showing the cold region near the valve orifice which comes in contact with cold Helium and the warm region which extends outside the cold mass of the cryomodule and is at room temperature. Delayed heat transfer between the hot and the cold regions trough convection of the Helium gas and conduction through the valve stalk leads to thermo-acoustic oscillations\cite{luck1992} and the resulting pressure waves resonate inside the closed space between the valve stem and the valve stalk. This mechanism of vibrations was first observed in the LCLS-II cryomodules while testing at Fermilab.\cite{Hansen2017} Following discussions with the Fermilab team, we inserted sleeves made of a cryogenic compatible PEEK plastic material on the stem to restrict the gas flow and suppress vibrations.\cite{Banerjee2018}

\begin{table*}[]
    \centering
    \begin{tabular}{|cc|ccc|ccc|}
        \hline
        Cavity & Stiffened & \multicolumn{3}{c|}{Peak Detuning (Hz)} & \multicolumn{3}{c|}{RMS Detuning (Hz)} \\
        \hline
        & & Original & JT and Precool & 5 K Adjust & Original & JT and Precool & 5 K Adjust \\
        & & & Static & Modified & & Static & Modified \\
        \hline
        1 & No  &  N/A  &  78   &  30  &  N/A  &  13.6  &  5.0 (4.7)\\
        2 & Yes &  18   &  N/A  &  25  &  4.4  &  N/A   &  4.6 (2.4)\\
        3 & No  &  280\textsuperscript{1}  &  100  &  50  &  11.2  &  20.8 &  10.7 (10.3)\\
        4 & Yes &  N/A  &  17   &  20  &  N/A  &  4.4   &  3.7 (2.4)\\
        5 & No  &  163\textsuperscript{1}  &  N/A  &  41  &  12.7  &  N/A  & 6.9 (6.3)\\
        6 & Yes &  30   &  N/A  &  18  &  5.0  &  N/A   &  3.2 (2.5)\\
        \hline
    \end{tabular}
    \caption{Microphonics measurements before and after cryogenic system modifications. RMS detuning is calculated from the detuning histograms except for the values in brackets which are obtained from the spectrum plots and are band limited to 200~Hz.}
    \label{tab:microcomp}
\end{table*}

Table~\ref{tab:microcomp} shows a summary of the microphonics measurements on all cavities in different configurations of the cryogenic system. The peak detuning on all unstiffened cavities(Fig.~\ref{fig:microimprov}) showed a significant reduction after the 5~K adjust valve was modified. However, we measured a new vibration line at 59 Hz which wasn't seen during our previous tests. The results from cross-correlation measurements of the microphonics detuning and accelerometer signals indicate that 59 Hz vibrations from an external source are being mechanically coupled into the cryomodule through the newly installed waveguides. The new source affected cavity 3 the most generating the highest peak detuning ($\sim 50\,\rm Hz$) among all others and this was an excellent candidate for testing the active compensation system.

Stiffened cavities did not show a significant reduction in peak microphonics detuning, with an increase being shown by cavities 2 and 4 even when the RMS diminished for cavity 4. Figure~\ref{fig:microimprov} shows results from microphonics measurements on these cavities. The histogram of detuning for cavity 2 and cavity 4 shows a flat top, indicating deviations from Gaussian white noise. The spectrum plot shows substantial vibration energy localized around 41~Hz and the 82~Hz corroborating this observation and the thermo-acoustic oscillations are indeed the reason as discussed earlier. The spectrum plots further indicate a reduction of the energy after valve modification in the same frequency bands, along with a net decrease in RMS detuning up to a vibration frequency of 200~Hz which is the limit of the data set. However, the width of the histograms also related to the RMS seems unchanged for cavity 4 and shows an increase in cavity 2 after valve modification seemingly contradicting the frequency domain observations. This apparent disagreement of the RMS detuning obtained from the histogram and the spectrum plots are listed in Tab.~\ref{tab:microcomp}. While the estimates agree for un-stiffened cavities, there is a significant difference for the stiffened cavities. The missing vibration energy could be accounted for by the excitation of high frequencies ($\gtrsim 200\,\rm{Hz}$), possibly the mechanical eigenmodes of the cavity along with measurement noise. Unfortunately the raw signals were not recorded during this experiment rendering us unable to analyze this in more detail.

\subsection{Active Compensation}
Passive measures of mitigating the vibration sources is the preferred method of reducing microphonics detuning. However active control is also necessary to temporarily restore operating gradient until the source is mitigated and to improve the margin of RF power consumption in the presence of existing microphonics detuning. We implemented the narrow band ANC algorithm in the Cornell Digital Low Level RF control system \cite{Neumann2013,Liepe2005} for this purpose. As described in Sec.~\ref{sec:anc}, the algorithm works separately for each frequency $\omega_m$ and requires two additional parameters, the gain $\mu_m$ and phase adaptation rate $\eta_m$ as described in Eq.~(\ref{eq:controller}) and Eq.~(\ref{eq:phiadj}) respectively. To determine an optimum setting, we start with small numbers for $\mu_m$ and $\eta_m$ until we start observing some effects on microphonics detuning, increasing $\mu_m$ until the feedback loop becomes unstable. We set $\mu_m$ to half of this maximum value to give us a suitable gain margin, and observe the controller phase $\phi_m$ as a function of time. We set $\eta_m$ so that the phase settles to the optimum value on average within a few minutes at the same time showing a noise level within $\pm 10^\circ$. Additionally, we nudged the frequency of compensation to find an optimum value within 0.1~Hz. This process is repeated for each frequency we want to compensate. The overall performance of resonance control depends on the quality factors of vibrations as explained in the previous section.

\begin{figure*}
    \centering
    \includegraphics[scale=0.5]{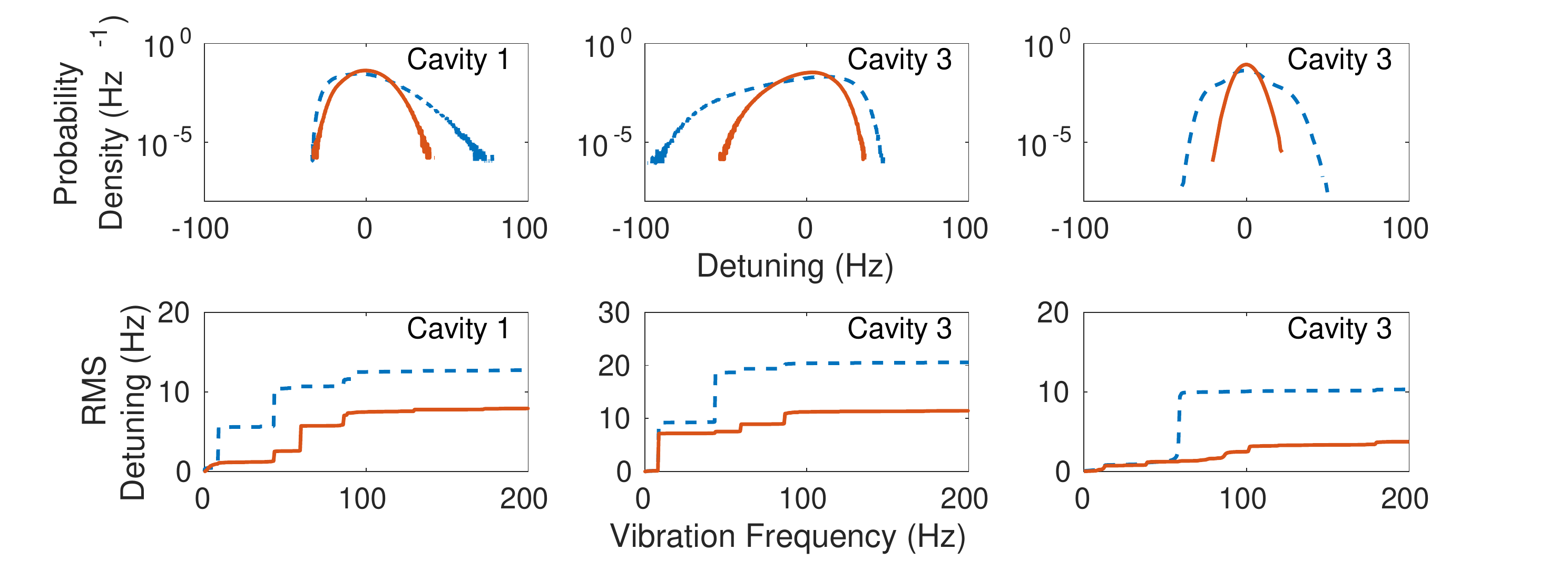}
    \caption{Effect of active microphonics compensation on two un-stiffened cavities of the main linac. The dashed lines represent data without active suppression while the bold lines show the performance with ANC turned on. The first two sets of data were taken before valve modification while the last data set was taken after.}
    \label{fig:ancresultsunstiff}
\end{figure*}

We have used the ANC algorithm during various stages of RF commissioning to attenuate microphonics and the results from un-stiffened cavities of the main linac are shown in Fig.~\ref{fig:ancresultsunstiff}. Before we modified the 5~K adjust valve, compensation was applied to 41~Hz and 8~Hz on un-stiffened cavities 1 and 3. The algorithm was successful in attenuating 41~Hz in both cavities 1 and 3 but was not effective on 8~Hz vibrations in cavity 3 as illustrated by the spectrum plots probably because the compensation frequency was not set correctly. These narrow band vibrations were a major contribution to microphonics detuning and their decrease also reduced the peak detuning. After we modified the 5~K adjust valve, we found the major source of microphonics detuning to be at 59 Hz. The ANC algorithm was successful in suppressing these vibrations in cavity 3. The attenuation of spectral lines are further validated by the RMS detuning as listed in Tab.~\ref{tab:ancresults}. The success of the algorithm indicates that those vibration lines were not in the vicinity of mechanical eigen modes of the tuner-cavity system which would have limited the effectiveness of the system as explained in Sec.~\ref{sec:anc}.

\begin{figure}
    \centering
    \includegraphics[scale=0.4]{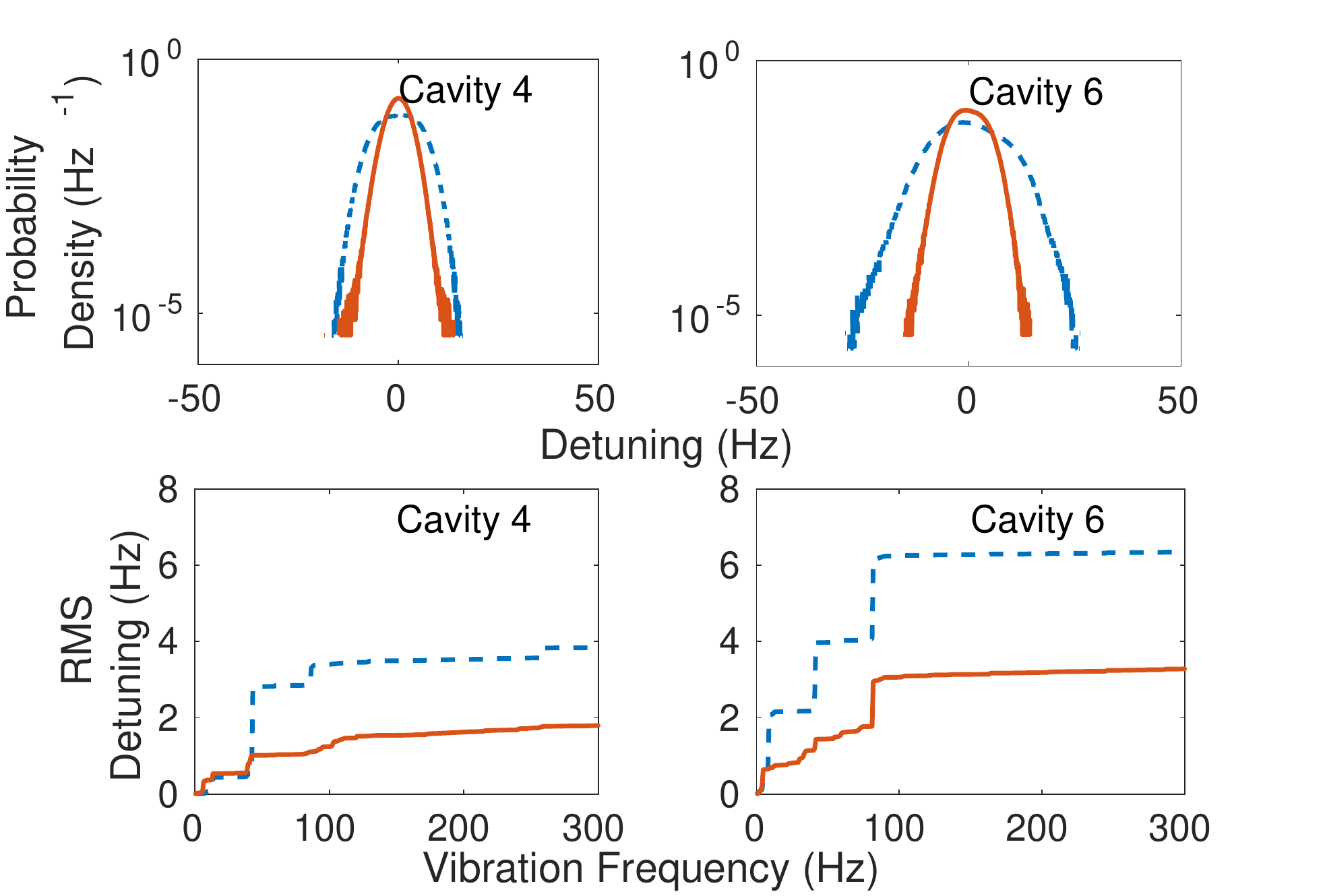}
    \caption{Effect of active microphonics compensation on two stiffened cavities of the main linac. The dashed lines represent data without active suppression while the bold lines show the performance with ANC turned on.}
    \label{fig:ancresultsstiff}
\end{figure}

\begin{figure}
    \centering
    \includegraphics[scale=0.4]{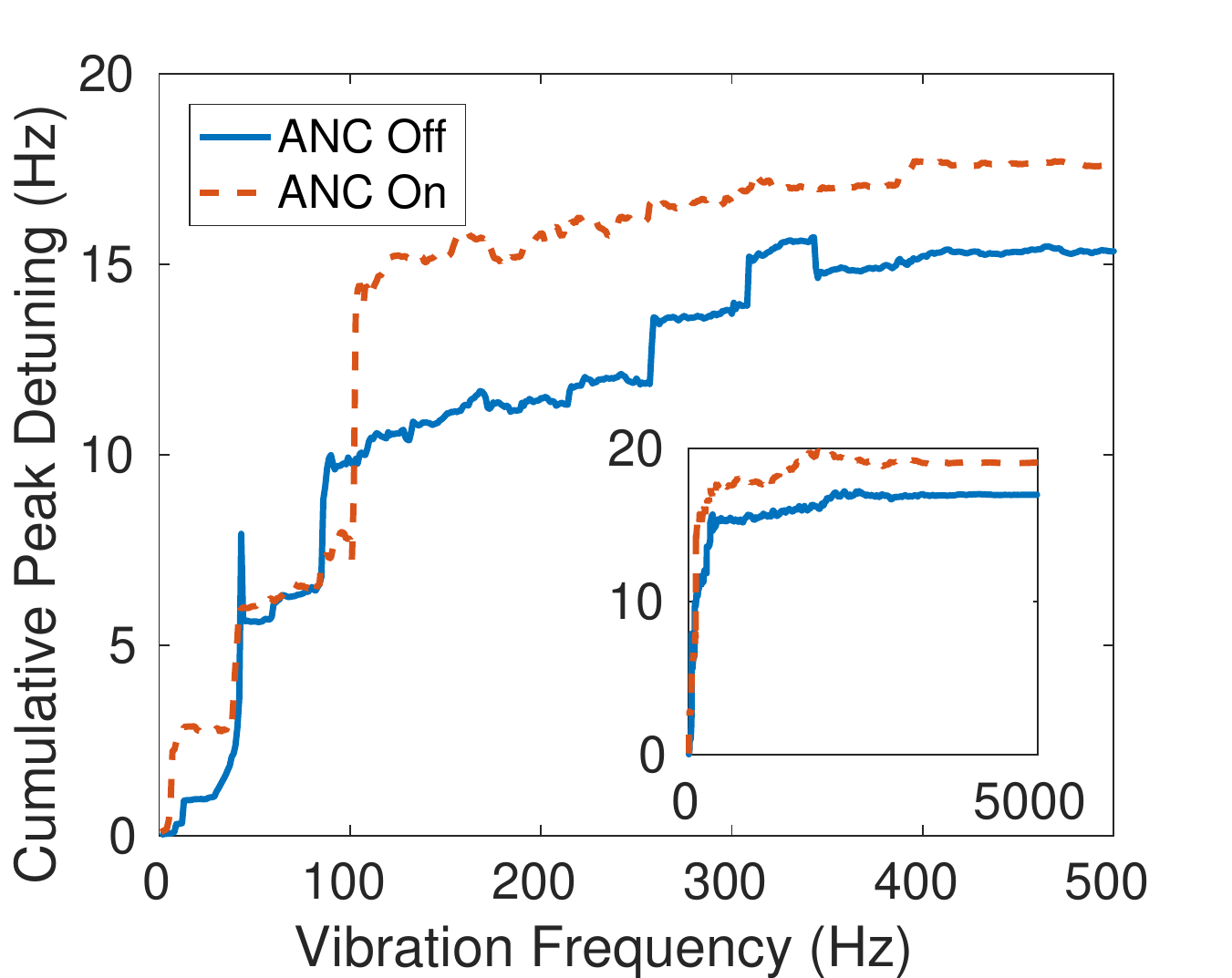}
    \caption{Cumulative peak detuning as a function of frequency for cavity 4 before valve modification showing the effect of active noise control system.}
    \label{fig:cav4peak}
\end{figure}

The results of using the system on stiffened cavities is shown in Fig.~\ref{fig:ancresultsstiff}. The algorithm was applied to cavities 4 and 6 for the frequencies 8~Hz and 41~Hz with additional attenuation of 82 Hz on cavity 6. While the ANC successfully reduced peak detuning from 30~Hz to 15~Hz in cavity 6, the measurements on cavity 4 indicate no reduction of peak detuning even though the RMS detuning is attenuated as seen from both the histogram and the spectrum plot. To understand which frequencies actually contribute to peak detuning, we Fourier transform the raw signal and zero all components beyond a certain vibration frequency and then find the peak detuning of the inverse transformed signal. Figure~\ref{fig:cav4peak} shows the cumulative peak detuning as a function of the vibration frequency threshold. When the ANC is off, 41~Hz and 82~Hz contribute most to the peak detuning as indicated by the large steps when we include these frequencies in the peak calculation. When we turn on compensation, the contribution from both these frequencies are reduced but a new mode at 102~Hz appears which accounts for almost half of the peak detuning but appears as a shallow step in the spectrum plot, illustrating it's transient nature. The ANC algorithm is thus well suited for compensating narrow band vibrations in both stiffened and un-stiffened cavities.

\begin{table*}[]
    \centering
    \begin{tabular}{|l|cc|cc| }
        \hline
        Run Description & \multicolumn{2}{c|}{Peak Detuning (Hz)} & \multicolumn{2}{c|}{RMS Detuning (Hz)} \\
        \hline
        & ANC Off & ANC On & ANC Off & ANC On \\
        \hline
        Cavity 1 with JT and precool static & 78 & 45 & 13.6 & 9.1\\
        Cavity 3 with JT and precool static & 100 & 57 & 20.8 & 11.7\\
        Cavity 3 with JT and precool static and 5~K adjust valve modified& 50 & 22 & 10.7 & 4.6\\
        Cavity 4 with JT and precool static & 17 & 19 & 4.4 & 2.4\\
        Cavity 6 in original configuration & 30 & 15 & 6.4 & 3.4\\
        \hline
    \end{tabular}
    \caption{Results of using the Active Noise Control system on various cavities during different stages of commissioning.}
    \label{tab:ancresults}
\end{table*}

The stability and robustness of the algorithm is demonstrated by comparatively long periods of stable operation with the same settings on different days. the observations shown in Tab.~\ref{tab:ancresults} are taken from data sets of at least 800~seconds measured for cavities inside a cryomodule connected to a production level cryogenic system unlike previous work primarily focused on test facilities. We achieved stable operations of over a few hours without spontaneous trips on all cavities with the ANC system active. We also successfully used it on cavity 3 during beam operations for the CBETA Fractional Arc Test which helped us achieve an energy gain of 8~MeV using a forward power below 5~kW which would not be possible without it. Once the settings were determined using the procedure explained earlier, resonance control was turn key with no tweaking required on subsequent days of operation which highlights the robustness of the system.

Lorentz Force Detuning (LFD) and mechanical coupling between different cavities in the cryomodule can be further sources of detuning which affect the operation of a resonance control system. The field dependence of LFD leads to decrease in the resonant frequency when the cavity field is ramped up. Large microphonics events generating a sudden increase in the resonance frequency of the cavity can lead to reduction in fields, LFD can in turn detune the cavity further in the positive direction amplifying the effect of the microphonics. Such an instability will lead to a catastrophic fall in cavity field and subsequent beam loss in an accelerator. However, in high Q machines the filling time of narrow bandwidth cavities can be sufficiently large, of the order of tens of milliseconds, slowing down the field decrease. This along with the presence of a high gain feedback loop on the field can be enough to avoid such an instability from developing. In all our operations till now, we have not used any feed-forward control of detuning and simple integral control of detuning has been enough to compensate for LFD when the field is ramped slowly. Further, the resonance control system of neighbouring cavities did not interact with each other during the course of normal operation since we have bellows mechanically isolating the cavities. This eliminates the need to account for such effects. The resonance control system described in this paper is thus a stable way of reducing peak detuning when mitigation of vibration sources is not an option.

\section{Conclusion}
The operation of SRF cavities with high $Q_L$ using solid state amplifiers of limited power present a significant constraint on the peak microphonics detuning which we can tolerate in order to maintain stable field. As a result, mitigation of vibration sources have become important for operations for a growing number of particle accelerators. Apart from passive measures, most cryomodules also incorporate fast tuners based on piezo-electric actuators which can be used for active resonance control such as in LCLS-II and XFEL. Assuming that the mechanics is adequately described by linear partial differential equations, we can model the tuner's response as a Linear Time Invariant system. This lets us encode the dynamics of the tuner in a transfer function $\tau(\omega)$, which expresses the amplitude and phase response of the cavity resonance frequency to sinusoidal excitations applied to the actuator at different frequencies. Measurements from 7-cell SRF cavities used in the main linac of the CBETA project indicate that the bandwidth of an active control system must be limited to below microphonics frequencies of 200~Hz beyond which mechanical eigen modes dominate the dynamics. The transfer function data is used in the design of the active microphonics control system.

We derived the narrow band Active Noise Control(ANC) algorithm starting from the assumption that microphonics detuning can be decomposed into a finite sum of sine waves. A sine wave of the same frequency applied to the actuator at the correct amplitude and phase should perfectly compensate for the vibrations. Using stochastic gradient descent, we derived an update relation which slowly changes phasors with components $I_m$ and $Q_m$ depending on the net microphonics detuning. These in turn modulate career signals at the microphonics frequencies $\omega_m$ applied to the actuator thus completing the ANC feedback controller. We further derived the frequency response of the ANC algorithm and established constraints on the adaptation rates $\mu_m$ and the controller phases $\phi_m$ to operate in the stable region, giving one concrete example from CBETA. Finally we propose a modification which automatically minimizes the mean square of compensated detuning by adapting the value of controller phase in response to changing tuner responses or vibration excitations. Using numerical simulations, we demonstrate the effectiveness of the modified ANC algorithm before implementing it on the main linac used in CBETA.

We applied various mitigation techniques on the main linac cryomodule of the CBETA project to reduce microphonics detuning. Passive measures included several modifications to the cryogenic system to damp thermo-acoustic oscillations and transients related to Helium flow and we achieved a reduction of peak detuning by at least a factor of 2. We further demonstrate the use of the active control system to achieve a stable reduction of microphonics typically by a factor of 2 without the need of detailed measurement of the transfer function. Future work will involve finding the remaining sources of vibrations and eliminating them, incorporating automatic frequency tracking in the ANC algorithm in order to make it more robust, and devising an algorithm to automatically configure the ANC on multiple frequencies during operation.

\section*{acknowledgment}
We would like to thank Vadim Veshcherevich and Roger Kaplan for extensive help in setting up the RF systems, Eric Smith, Dan Sabol and Colby Shore for setting up the cryogenic systems including procuring the plastic sleeves and modifying the valve stem, Adam Bartnik and Colwyn Gulliford for helping with operations. We are grateful to Warren Schappert and Ben Hansen for sharing the experience of commissioning the LCLS-II cryomodules at Fermilab which prompted us to look for thermo-acoustic oscillations in our cryogenic valves. We are indebted to Claudio Rivetta and Douglas MacMartin for providing insight on theoretical aspects of our controller. This work was supported by the New York State Energy Research and Development Authority. CLASSE facilities are operated with major support from the National Science Foundation.


\begin{thebibliography}{10}

\bibitem{Hasan2014}
Hasan~S. Padamsee.
\newblock Superconducting radio-frequency cavities.
\newblock {\em Annual Review of Nuclear and Particle Science}, 64(1):175--196,
  2014.

\bibitem{Mastorides2010}
T.~Mastorides, C.~Rivetta, J.~D. Fox, D.~Van Winkle, and P.~Baudrenghien.
\newblock Rf system models for the cern large hadron collider with application
  to longitudinal dynamics.
\newblock {\em Phys. Rev. ST Accel. Beams}, 13:102801, Oct 2010.

\bibitem{Belomestnykh2001}
S.~Belomestnykh and H.~Padamsee.
\newblock {Performance of the CESR Superconducting RF System and Future Plans}.
\newblock {\em SRF Workshop}, 2001.

\bibitem{Rose2011}
B.~Holub Y. Kawashima H. Ma N. Towne M.~Yeddulla J.~Rose, W.~Gash.
\newblock Nsls-ii rf systems.
\newblock {\em Proceedings, PAC}, 2011.
\newblock FROBS4.

\bibitem{Doolittle2015}
A~Ratti C Serrano R Bachimanchi C Hovater S Babel B Hong D Van Winkle B Chase E
  Cullerton P~Varghese L~Doolittle, G~Huang.
\newblock The lcls-ii llrf system.
\newblock {\em Proceedings, IPAC}, 2015.
\newblock MOPWI021.

\bibitem{Branlard2013}
Ł~Butkowski H Schlarb J Sekutowicz W Cichalewski A Piotrowski K Przygoda W
  Jałmu˙zna J~Szewi´nski J~Branlard, V~Ayvazyan.
\newblock Llrf system design and performance for xfel cryomodules continuous
  wave operation.
\newblock {\em SRF Workshop}, 2013.
\newblock THP086.

\bibitem{Hoffstaetter2017}
G.~H. Hoffstaetter et~al.
\newblock {CBETA Design Report, Cornell-BNL ERL Test Accelerator}.
\newblock {\em arXiv}, 2017.
\newblock 1706.04245.

\bibitem{AboBakr2018}
Michael Abo-Bakr et~al.
\newblock {Status Report of the Berlin Energy Recovery Linac Project
  BERLinPro}.
\newblock In {\em Proc. 9th International Particle Accelerator Conference
  (IPAC'18), Vancouver, BC, Canada, April 29-May 4, 2018}, number~9 in
  International Particle Accelerator Conference, page THPMF034, Geneva,
  Switzerland, June 2018. JACoW Publishing.

\bibitem{Liepe2003}
M.~Liepe and S.~Belomestnykh.
\newblock {RF Parameter and Field Stability Requirements for the Cornell ERL
  Prototype}.
\newblock In {\em {Particle accelerator. Proceedings, Conference, PAC 2003,
  Portland, USA, May 12-16, 2003}}, volume C030512, page 1329, 2003.

\bibitem{Posen2012PRAB}
Sam Posen and Matthias Liepe.
\newblock {Mechanical optimization of superconducting cavities in continuous
  wave operation}.
\newblock {\em Phys. Rev. ST Accel. Beams}, 15:022002, 2012.

\bibitem{Eichhorn2014}
Ralf Eichhorn et~al.
\newblock {Cornell's Main Linac Cryomodule for the Energy Recovery Linac
  Project}.
\newblock In {\em {Proceedings, 5th International Particle Accelerator
  Conference (IPAC 2014): Dresden, Germany, June 15-20, 2014}}, number~5 in
  International Particle Accelerator Conference, page WEPRI061, Geneva,
  Switzerland, June 2014. JACoW Publishing.

\bibitem{Posen2012ipac}
S.~Posen and M.~Liepe.
\newblock {Measurement of the Mechanical Properties of Superconducting Cavities
  During Operation}.
\newblock In {\em {Proceedings, 3rd International Conference on Particle
  accelerator (IPAC 2012): New Orleans, USA, May 2-25, 2012}}, volume C1205201,
  pages 2399--2401, 2012.

\bibitem{Pischalnikov2015}
Yuriy Pischalnikov, Butch Hartman, Jeremiah Holzbauer, Warren Schappert, Samuel
  Smith, and Jae-Chul Yun.
\newblock {Reliability of the LCLS II SRF Cavity Tuner}.
\newblock In {\em {Proceedings, 17th International Conference on RF
  Superconductivity (SRF2015): Whistler, Canada, September 13-18, 2015}}, page
  THPB065, 2015.

\bibitem{Cichalewski2015}
Wojciech Cichalewski, Julien Branlard, Andrzej Napieralski, and Christian
  Schmidt.
\newblock {European XFEL Cavities Piezoelectric Tuners Control Range
  Optimization}.
\newblock In {\em {Proceedings, 15th International Conference on Accelerator
  and Large Experimental Physics Control Systems (ICALEPCS 2015): Melbourne,
  Australia, October 17-23, 2015}}, page MOPGF079, 2015.

\bibitem{Elliott2015}
S.~J. Elliott, M.~Ghandchi Tehrani, and R.~S. Langley.
\newblock Nonlinear damping and quasi-linear modelling.
\newblock {\em Philosophical Transactions of the Royal Society of London A:
  Mathematical, Physical and Engineering Sciences}, 373(2051), 2015.

\bibitem{Conway2010}
Zachary Conway and Matthias Liepe.
\newblock {Fast Piezoelectric Actuator Control of Microphonics in the CW
  Cornell ERL Injector Cryomodule}.
\newblock In {\em {Particle accelerator. Proceedings, 23rd Conference, PAC'09,
  Vancouver, Canada, May 4-8, 2009}}, page TU5PFP043, 2010.

\bibitem{Banerjee2017}
N.~Banerjee et~al.
\newblock {M}icrophonics {S}tudies of the {CBETA} {L}inac {C}ryomodules.
\newblock In {\em Proc. of International Particle Accelerator Conference
  (IPAC'17), Copenhagen, Denmark, 14â19 May, 2017}, number~8 in
  International Particle Accelerator Conference, pages 1138--1141, Geneva,
  Switzerland, May 2017. JACoW.
\newblock https://doi.org/10.18429/JACoW-IPAC2017-MOPVA122.

\bibitem{Neumann2010}
A.~Neumann, W.~Anders, O.~Kugeler, and J.~Knobloch.
\newblock Analysis and active compensation of microphonics in continuous wave
  narrow-bandwidth superconducting cavities.
\newblock {\em Phys. Rev. ST Accel. Beams}, 13:082001, Aug 2010.

\bibitem{EinsteinCurtis2017}
Joshua Einstein-Curtis.
\newblock Microphonics and active compensation.
\newblock Low Level Radio Frequency Workshop, 2017.

\bibitem{Rybaniec2017}
R.~Rybaniec, K.~Przygoda, W.~Cichalewski, V.~Ayvazyan, J.~Branlard, Ł.
  Butkowski, S.~Pfeiffer, C.~Schmidt, H.~Schlarb, and J.~Sekutowicz.
\newblock Fpga-based rf and piezocontrollers for srf cavities in cw mode.
\newblock {\em IEEE Transactions on Nuclear Science}, 64(6):1382--1388, June
  2017.

\bibitem{Kuo1997}
S.~M. Kuo and D.~R. Morgan.
\newblock Active noise control: a tutorial review.
\newblock {\em Proceedings of the IEEE}, 87(6):943--973, Jun 1999.

\bibitem{Hoffstaetter2017Linac}
Georg Hoffstaetter et~al.
\newblock {CBETA: The Cornell/BNL 4-Turn ERL with FFAG Return Arcs for eRHIC
  Prototyping}.
\newblock In {\em {Proceedings, 28th International Linear Accelerator
  Conference (LINAC16): East Lansing, Michigan, September 25-30, 2016}}, page
  TUOP02, 2017.

\bibitem{Trbojevic2017Ipac}
Dejan Trbojevic et~al.
\newblock {CBETA - Cornell University Brookhaven National Laboratory Electron
  Energy Recovery Test Accelerator}.
\newblock In {\em {Proceedings, 8th International Particle Accelerator
  Conference (IPAC 2017): Copenhagen, Denmark, May 14-19, 2017}}, page TUOCB3,
  2017.

\bibitem{Hoffstaetter2012}
M.~Liepe, G.~H. Hoffstaetter, S.~Posen, P.~Quigley, and V.~Veshcherevich.
\newblock {High Current Operation of the Cornell ERL Superconducting RF
  Injector Cryomodule}.
\newblock In {\em {Proceedings, 3rd International Conference on Particle
  accelerator (IPAC 2012): New Orleans, USA, May 2-25, 2012}}, volume C1205201,
  pages 2378--2380, 2012.

\bibitem{Hartill2011}
M.~Liepe, D.~L. Hartill, G.~H. Hoffstaetter, S.~Posen, P.~Quigley, and
  V.~Veshcherevich.
\newblock {Experience with the Cornell ERL Injector SRF Cryomodule during High
  Beam Current Operation}.
\newblock In {\em {Particle accelerator. Proceedings, 2nd International
  Conference, IPAC 2011, San Sebastian, Spain, September 4-9, 2011}}, volume
  C110904, pages 35--37, 2011.

\bibitem{Liepe2010}
Matthias Liepe, Sergey Belomestnykh, Eric Chojnacki, Zachary Conway, Valeri
  Medjidzade, Hasan Padamsee, Peter Quigley, James Sears, Valery Shemelin, and
  Vadim Veshcherevich.
\newblock {SRF Experience with the Cornell High-Current ERL Injector
  Prototype}.
\newblock In {\em {Particle accelerator. Proceedings, 23rd Conference, PAC'09,
  Vancouver, Canada, May 4-8, 2009}}, page TU3RAI01, 2010.

\bibitem{Eichhorn2015Cryo}
R.~Eichhorn et~al.
\newblock {The Cornell Main Linac Cryomodule: A Full Scale, High Q Accelerator
  Module for cw Application}.
\newblock In {\em {Proceedings, International Cryogenic Engineering Conference
  and International Cryogenic Materials Conference 2014 (ICEC 25 – ICMC
  2014): Enschede, The Netherlands, July 7-11, 2014}}, volume~67, pages
  785--790, 2015.

\bibitem{Furuta2016}
Fumio Furuta et~al.
\newblock {ERL Main Linac Cryomodule Cavity Performance and Effect of Thermal
  Cycling}.
\newblock In {\em {Proceedings, 7th International Particle Accelerator
  Conference (IPAC 2016): Busan, Korea, May 8-13, 2016}}, page WEPMR022, 2016.

\bibitem{He2012}
Y.~He, G.~H. Hoffstaetter, M.~Liepe, M.~Tigner, and E.~N. Smith.
\newblock {Cryogenic Distribution System for the Proposed Cornell ERL Main
  Linac}.
\newblock In {\em {Proceedings, 3rd International Conference on Particle
  accelerator (IPAC 2012): New Orleans, USA, May 2-25, 2012}}, volume C1205201,
  pages 619--621, 2012.

\bibitem{luck1992}
H.~Luck and Ch. Trepp.
\newblock Thermoacoustic oscillations in cryogenics. part 1: basic theory and
  experimental verification.
\newblock {\em Cryogenics}, 32(8):690 -- 697, 1992.

\bibitem{Hansen2017}
B~J Hansen, O~Al Atassi, R~Bossert, J~Einstein-Curtis, J~Holzbauer, W~Hughes,
  J~Hurd, J~Kaluzny, A~Klebaner, J~Makara, Y~Pischalnikov, W~Schappert,
  R~Stanek, J~Theilacker, R~Wang, and M~J White.
\newblock Effects of thermal acoustic oscillations on lcls-ii cryomodule
  testing.
\newblock {\em IOP Conference Series: Materials Science and Engineering},
  278(1):012188, 2017.

\bibitem{Banerjee2018}
N~Banerjee, J~Dobbins, F~Furuta, G~Hoffstaetter, R~Kaplan, M~Liepe, P~Quigley,
  E~Smith, and V~Veshcherevich.
\newblock Microphonics suppression in the cbeta linac cryomodules.
\newblock {\em Journal of Physics: Conference Series}, 1067(8):082004, 2018.

\bibitem{Neumann2013}
A.~Neumann, W.~Anders, R.~Goergen, J.~Knobloch, O.~Kugeler, S.~Belomestnykh,
  J.~Dobbins, R.~Kaplan, M.~Liepe, and C.~Strohman.
\newblock Cw measurements of cornell llrf system at hobicat.
\newblock In {\em Proceedings, 15th Superconducting Radio Frequency Workshop
  (SRF '2011), Chicago, USA, July 25-29, 2011}, number~15 in Superconducting
  Radio Frequency Workshop, page MOPO67, Geneva, Switzerland, July 2011. JACoW
  Publishing.

\bibitem{Liepe2005}
M.~Liepe, S.~Belomestnykh, J.~Dobbins, R.~Kaplan, C.~Strohman, and B.~Stuhl.
\newblock Experience with the new digital rf control system at the cesr storage
  ring.
\newblock In {\em Proceedings of the 2005 Particle Accelerator Conference},
  pages 2592--2594, May 2005.

\end{thebibliography}
\end{document}